\documentclass[letter,apj]{emulateapj-rtx4}
\usepackage{graphicx}
\usepackage{enumerate}
\usepackage{amssymb, amsmath}
\usepackage{natbib}
\usepackage{color}
\usepackage{ulem}

\newcommand{\princeton}{1}
\newcommand{\tokodai}{2}
\newcommand{\goddard}{3}
\newcommand{\mpia}{4}
\newcommand{\oklahoma}{5}
\newcommand{\physmath}{6}
\newcommand{\charleston}{7}
\newcommand{\kyoto}{8}
\newcommand{\harvard}{9}
\newcommand{\madrid}{10}
\newcommand{\pannekoek}{11}
\newcommand{\subaru}{12}
\newcommand{\naoj}{13}
\newcommand{\fizeau}{14}
\newcommand{\toronto}{15}
\newcommand{\sternwarte}{16}
\newcommand{\eureka}{17}
\newcommand{\tokyo}{18}
\newcommand{\ifahawaii}{19}
\newcommand{\hiroshima}{20}
\newcommand{\jpl}{21}
\newcommand{\sokendai}{22}
\newcommand{\sinica}{23}
\newcommand{\sapporo}{24}
\newcommand{\sendai}{25}

\begin{document}

\title{The Moving Group Targets of the SEEDS High-Contrast Imaging Survey of Exoplanets and Disks: Results and Observations from the First Three Years}
\author{Timothy D.~Brandt\altaffilmark{\princeton}, 
Masayuki Kuzuhara\altaffilmark{\tokodai},
Michael W.~McElwain\altaffilmark{\goddard}, 
Joshua E.~Schlieder\altaffilmark{\mpia},
John~P. Wisniewski\altaffilmark{\oklahoma}, 
Edwin L.~Turner\altaffilmark{\princeton,\physmath},
   J. Carson\altaffilmark{\charleston,\mpia},  
   T. Matsuo\altaffilmark{\kyoto},   
   B. Biller\altaffilmark{\mpia},
   M. Bonnefoy\altaffilmark{\mpia},
   C. Dressing\altaffilmark{\harvard},  
   M. Janson\altaffilmark{\princeton},   
   G.~R. Knapp\altaffilmark{\princeton}, 
   A. Moro-Mart\'in\altaffilmark{\madrid},  
   C. Thalmann\altaffilmark{\pannekoek},  
   T. Kudo\altaffilmark{\subaru},   
   N. Kusakabe\altaffilmark{\naoj},   
   J. Hashimoto\altaffilmark{\naoj,\oklahoma}, 
   L. Abe\altaffilmark{\fizeau},   
   W. Brandner\altaffilmark{\mpia},     
   T. Currie\altaffilmark{\toronto},
   S. Egner\altaffilmark{\subaru},   
   M. Feldt\altaffilmark{\mpia},  
   T. Golota\altaffilmark{\subaru},   
   M. Goto\altaffilmark{\sternwarte},  
   C.~A. Grady\altaffilmark{\goddard,\eureka}, 
   O. Guyon\altaffilmark{\subaru},  
   Y. Hayano\altaffilmark{\subaru},  
   M. Hayashi\altaffilmark{\tokyo},  
   S. Hayashi\altaffilmark{\subaru}, 
   T. Henning\altaffilmark{\mpia},   
   K.~W. Hodapp\altaffilmark{\ifahawaii},  
   M. Ishii\altaffilmark{\subaru}, 
   M. Iye\altaffilmark{\naoj},    
   R. Kandori\altaffilmark{\naoj},  
   J. Kwon\altaffilmark{\naoj,\sokendai},
   K. Mede\altaffilmark{\tokyo},
   S. Miyama\altaffilmark{\hiroshima},   
   J.-I. Morino\altaffilmark{\naoj},  
   T. Nishimura\altaffilmark{\subaru},
   T.-S. Pyo\altaffilmark{\subaru}, 
   E. Serabyn\altaffilmark{\jpl},   
   T. Suenaga\altaffilmark{\sokendai},
   H. Suto\altaffilmark{\naoj},    
   R. Suzuki\altaffilmark{\naoj},   
   M. Takami\altaffilmark{\sinica},  
   Y. Takahashi\altaffilmark{\tokyo},
   N. Takato\altaffilmark{\subaru}, 
   H. Terada\altaffilmark{\subaru},  
   D. Tomono\altaffilmark{\subaru},  
   M. Watanabe\altaffilmark{\sapporo},  
   T. Yamada\altaffilmark{\sendai},   
   H. Takami\altaffilmark{\subaru},  
   T. Usuda\altaffilmark{\subaru}, 
   M. Tamura\altaffilmark{\naoj,\tokyo}  
}

\altaffiltext{*}{Based on data collected at Subaru Telescope, which
   is operated by the National Astronomical Observatory of Japan.}

\altaffiltext{\princeton}{Department of Astrophysical Sciences, Princeton University, Princeton, NJ, USA.}
\altaffiltext{\tokodai}{Tokyo Institue of Technology, Tokyo, Japan.}
\altaffiltext{\goddard}{Exoplanets and Stellar Astrophysics Laboratory, NASA Goddard Space Flight Center, Greenbelt, MD, USA.}
\altaffiltext{\mpia}{Max Planck Institute for Astronomy, Heidelberg, Germany.}
\altaffiltext{\oklahoma}{HL Dodge Department of Physics and Astronomy, University of Oklahoma, Norman, OK, USA.}
\altaffiltext{\physmath}{Kavli Institute for the Physics and Mathematics of the Universe (WPI), Todai Institutes for Advanced Study, University of Tokyo, Japan.}
\altaffiltext{\charleston}{College of Charleston, Charleston, SC, USA.}
\altaffiltext{\kyoto}{Department of Astronomy, Kyoto University, Kyoto, Japan.}
\altaffiltext{\harvard}{Harvard-Smithsonian Center for Astrophysics, Cambridge, MA, USA.}
\altaffiltext{\madrid}{Department of Astrophysics, CAB - CSIC/INTA, Madrid, Spain.}
\altaffiltext{\pannekoek}{Astronomical Institute Anton Pannekoek, 
    University of Amsterdam, Amsterdam, The Netherlands.}
\altaffiltext{\subaru}{Subaru Telescope, Hilo, Hawai`i, USA.}
\altaffiltext{\naoj}{National Astronomical Observatory of Japan, Tokyo, Japan.}

\altaffiltext{\fizeau}{Laboratoire Hippolyte Fizeau, Nice, France.}
\altaffiltext{\toronto}{University of Toronto, Toronto, Canada.}
\altaffiltext{\sternwarte}{Universit\"ats-Sternwarte M\"unchen, Munich, Germany.}
\altaffiltext{\eureka}{Eureka Scientific, Oakland, CA, USA.}

\altaffiltext{\tokyo}{University of Tokyo, Tokyo, Japan.}
\altaffiltext{\ifahawaii}{Institute for Astronomy, University of Hawai`i, Hilo, Hawai`i, USA.}
\altaffiltext{\hiroshima}{Hiroshima University, Higashi-Hiroshima, Japan.}
\altaffiltext{\jpl}{Jet Propulsion Laboratory, California Institute of Technology, Pasadena, CA, USA.}

\altaffiltext{\sokendai}{Department of Astronomical Science, Graduate University for Advanced Studies, Tokyo, Japan.}
\altaffiltext{\sinica}{Institute of Astronomy and Astrophysics, Academia Sinica, Taipei, Taiwan.}
\altaffiltext{\sapporo}{Department of Cosmosciences, Hokkaido University, Sapporo, Japan.}
\altaffiltext{\sendai}{Astronomical Institute, Tohoku University, Sendai, Japan.}

\begin{abstract}
We present results from the first three years of observations of moving group targets in the SEEDS high-contrast imaging survey of exoplanets and disks using the Subaru telescope.  We achieve typical contrasts of $\sim$10$^5$ at 1$''$ and $\sim$10$^6$ beyond 2$''$ around 63 proposed members of nearby kinematic moving groups.  We review each of the kinematic associations to which our targets belong, concluding that five, $\beta$ Pictoris ($\sim$20 Myr), AB Doradus ($\sim$100 Myr), Columba ($\sim$30 Myr), Tucana-Horogium ($\sim$30 Myr), and TW Hydrae ($\sim$10 Myr), are sufficiently well-defined to constrain the ages of individual targets.  Somewhat less than half of our targets are high-probability members of one of these moving groups.  For all of our targets, we combine proposed moving group membership with other age indicators where available, including Ca\,{\sc ii} HK emission, X-ray activity, and rotation period, to produce a posterior probability distribution of age.  SEEDS observations discovered a substellar companion to one of our targets, $\kappa$ And, a late B star.  We do not detect any other substellar companions, but do find seven new close binary systems, of which one still needs to be confirmed.  A detailed analysis of the statistics of this sample, and of the companion mass constraints given our age probability distributions and exoplanet cooling models, will be presented in a forthcoming paper.
\end{abstract}

\section{Introduction} \label{sec:intro}
More than 850 exoplanets are now known to orbit other stars.  Most were identified with indirect detection techniques, but exoplanets have now been imaged around several young, nearby stars.  Direct imaging is the primary technique used to probe the frequency of giant exoplanets at separations similar to the outer solar system ($\sim$4-40 AU).  It is already providing important constraints on planetary formation mechanisms, complementing the well-characterized distribution and frequency of planets at separations similar to the inner solar system (e.g., \citealt{Cumming+Butler+Marcy+etal_2008}, \citealt{Howard+Marcy+Johnson+etal_2010}).  By measuring its emission spectrum, direct imaging constrains an exoplanetary atmosphere's temperature, composition, and dynamics.  Direct imaging of exoplanets of known age can also break the mass-age-luminosity degeneracy in exoplanet cooling models.

Several surveys have set out to directly image exoplanets around nearby stars.  The direct imaging of exoplanets is challenging observationally, due to their high contrast ($\gtrsim$10$^{4}$) and small separations from the host star ($\lesssim$ 1$\arcsec$).  These observational requirements are mitigated by targeting young, nearby systems (See \citealt{Oppenheimer+Hinkley_2009}).  Young exoplanets cool rapidly as they radiate away their residual heat of formation, quickly falling below the detectability limits of even the largest ground-based telescopes equipped with high-contrast instrumentation.  This sub-stellar evolution is similar to that of brown dwarfs, but is distinct from stellar evolution (e.g., \citealt{Burrows+Marley+Hubbard+etal_1997}, \citealt{Chabrier+Baraffe+Allard+etal_2000}).  Nearby stars are important because the angular resolution is set by Earth's atmosphere and the optical system; nearer stars can therefore probe smaller physical separations.  Unfortunately, the vast majority of observations remain null detections:  massive exoplanets and brown dwarfs at large separations appear to be rare (c.f., \citealt{McCarthy+Zuckerman_2004}, \citealt{Masciadri+Mundt+Henning+etal_2005}, \citealt{Carson+Eikenberry+Smith+etal_2006}, \citealt{Lafreniere+Doyon+Marois+etal_2007}, \citealt{Biller+Close+Maciadri+etal_2007}, \citealt{Metchev+Hillenbrand_2009}, \citealt{Janson+Bonavita+Klahr+etal_2011}, \citealt{Vigan+Patience+Marois+etal_2012}).  In order to properly interpret these results, however, uncertainties in stellar ages must be taken into account.

We report the strategy and results from the first three years of the `Moving Groups' subcategory of the Strategic Exploration of Exoplanets and Disks with Subaru (SEEDS) survey \citep{Tamura_2009}.  A Moving Group is a collection of stars that share a common age, metallicity, and space motion due to formation in the same event.  Nearby Moving Group stars are particularly promising targets for direct imaging planet searches due to their proximity and well defined youthful ages.  The SEEDS survey itself is briefly described in \S~\ref{sec:seeds_overview}.  The architecture and target selection strategy of the SEEDS Moving Groups sub-category is discussed in \S~\ref{sec:moving_groups}.  This section also includes a review of each of the moving groups that were drawn upon for the target sample, as well as the age indicators used for the targets.  In \S~\ref{sec:otherindicators}, individual stellar age indicators are described in the context of how they were implemented to constrain the ages of the target sample.  Section \ref{sec:BayesAges} describes the Bayesian approach to assign statistically significant stellar ages for the target sample.  The observations and details regarding individual stars are discussed in \S~\ref{sec:obs_details}.  The data reduction details are summarized in \S~\ref{sec:data_reduction}, and a discussion of the Moving Group sample sensitivity is presented in \S~\ref{sec:discussion}.  The concluding remarks are presented in \S~\ref{sec:conclusions}.  

\section{The SEEDS Survey} 
\label{sec:seeds_overview}

The SEEDS survey is the most ambitious high-contrast imaging survey to date.  This survey is being carried out with a suite of high-contrast instrumentation at the Subaru Telescope, including a second generation adaptive optics (AO) system with 188 actuators \citep[AO188, ][]{Hayano+Takami+Guyon+etal_2008} and a dedicated differential imaging instrument called HiCIAO \citep{Suzuki+Kudo+Hashimoto+etal_2010}.  SEEDS is now $\sim$$2/3$ complete, and will ultimately observe $\sim$500 stars to search for exoplanets and disks with direct imaging.  

The SEEDS survey is organized into two separate classes: planets and disks.  Each of SEEDS' target classes, planets and disks, is further subdivided into categories, including nearby stars, moving groups (MG; this work), debris disks (Janson et al. 2013, submitted), young stellar objects (containing the protoplanetary and transitional disks), and open clusters (Yamamoto et al. 2013, accepted).  The nearby stars category is further separated into sub-categories that include high mass stars (Carson et al. 2013, in preparation), M-dwarfs, white dwarfs, chromospherically active stars, stars with kinematic properties suggestive of youth, and stars with known radial velocity planets (e.g., \citealt{Narita+Kudo+Bergfors+etal_2010}, \citealt{Narita+Takahashi+Kuzuhara+etal_2012}).

HiCIAO offers several observing modes, including polarized differential imaging (PDI), simultaneous imaging at different wavelengths (spectral differential imaging, or SDI), and simple direct imaging (DI, or angular differential imaging [ADI] when used with the image rotator off and the pupil rotation angle fixed on the detector).  Young disks, with plentiful scattering by small grains, are typically observed in polarized light (PDI mode).  PDI obtains simultaneous measurements of perpendicular polarization states; the two images are later subtracted to remove unpolarized light \citep{Kuhn+Potter+Parise_2001}.  SEEDS implements the double difference technique that subtracts a similar polarization scene modulated by 90$\degr$, effectively removing the non common path errors between the channels (e.g., \citealt{Hinkley+Oppenheimer+Soummer+etal_2009}).  Older debris disks have much weaker polarized scattering; only their total scattered intensity is typically observed.  All stars without disks predicted from infrared excesses are observed only in total intensity (DI), and the data are processed using ADI.  

Early survey highlights include three directly detected substellar companions, GJ 758 B \citep{Thalmann+Carson+Janson+etal_2009, Janson+Carson+Thalmann+etal_2011}, $\kappa$ Andromedae b \citep{Carson+Thalmann+Janson+etal_2013}, and GJ 504 b \citep{Kuzuhara+Tamura+Kudo+etal_2013, Janson+Brandt+Kuzuhara+etal_2013}.  In addition, there has been a plethora of papers that investigate circumstellar disk properties in the protoplanetary \citep{Hashimoto+Tamura+Muto+etal_2011, Kusakabe+Grady+Sitko+etal_2012}, transitional \citep{Thalmann+Grady+Goto+etal_2010, Muto+Grady+Hashimoto+etal_2012, Hashimoto+Dong+Kudo+etal_2012, Dong+Hashimoto+Rafikov+etal_2012, Mayama+Hashimoto+Muto+etal_2012, Tanii+Itoh+Kudo+etal_2012, Grady+Muto+Hashimoto+etal_2013, Follette+Tamura+Hashimoto+etal_2013}, and debris \citep{Thalmann+Janson+Buenzli+etal_2011, Thalmann+Janson+Buenzli+etal_2013} phases of evolution.  These include some of the first near-IR images of protoplanetary and transitional disks, including hints of substellar companions from disk structure, and characterizations of debris disks believed to be generated by the destruction of planetesimals.  

The goal of the SEEDS survey is to provide observational constraints on all stages of exoplanet formation and evolution, from protoplanetary and transitional disks to older, disk-free systems.  The survey therefore targets a wide range of host stars.  Unfortunately, many of the SEEDS targets, while they do show indicators of youth, lack well-determined ages.  This leads to large uncertainties when converting exoplanet luminosities into masses using theoretical cooling models (e.g., \citealt{Burrows+Marley+Hubbard+etal_1997}, \citealt{Baraffe+Chabrier+Barman+etal_2003}, \citealt{Marley+Fortney+Hubickyj+etal_2007}, \citealt{Spiegel+Burrows_2012}).  The MG category is designed to overcome this problem by observing nearby stars reliably associated with kinematic moving groups $\sim$10--500 Myr old.  

Because of their distances and ages, the SEEDS MG sample includes some of the most promising targets in the sky for the direct detection of exoplanets.  Many of these targets have been observed by other previous and ongoing surveys, and we make use of the publicly available data in our analysis, primarily as a means of identifying background stars in the field of view by confirming they do not share common proper motion with the target star (See \S~\ref{sec:obs_details}).

\section{SEEDS Moving Groups} 
\label{sec:moving_groups}

Many of the youngest stars near the Sun are members of moving groups, loose associations of stars defined by their common Galactic kinematics and ages (See reviews by \citealt{Zuckerman+Song_2004, Torres+Quast+Melo+etal_2008}). Some moving groups have been kinematically and chemically associated with nearby clusters, linking them to recent episodes of star formation near the Sun \citep{Mamajek+Feigelson_2001, Ortega+delaReza+Jilinski+etal_2002, Fernandez+Figueras+Torra_2008, Barenfeld+Bubar+Mamjek+etal_2013, DeSilva+DOrazi+Melo+etal_2013}. Moving groups have members within the solar neighborhood ($\lesssim$100 pc) and ages $\sim$10-500 Myr.  If a proposed moving group is real, and not a dynamical stream (see the following subsections), the true members are coeval. Group ages are determined using many methods based on both individual proposed members and the group as an aggregate.  These include:  HR diagrams, isochrone fitting, lithium depletion, chromospheric and coronal emission, rotation, and the kinematic trace back of the group members to the most compact volume in space where they were formed coevally.  The likelihood that a star is a true moving group member depends on both its kinematics and youth indicators.  The targets for the SEEDS Moving Groups category are proposed members of the nearby, young kinematic moving groups AB Doradus, $\beta$ Pictoris, Castor, Columba, Hercules-Lyra, the IC 2391 supercluster, the Local Association, Tucana-Horologium, TW Hydrae, and Ursa Major/Sirius.  We briefly summarize each of these associations in the following subsections.  

\subsection{The AB Doradus Moving Group}
\label{subsec:ABDor}

\cite{Torres+Quast+deLaReza+etal_2003} and \cite{Zuckerman+Song+Bessel_2004} independently proposed the AB Doradus moving group via searches for stars with common kinematics and ages in publicly available catalogs.  AB Dor has one of the largest proposed membership samples of any moving group---\cite{Torres+Quast+Melo+etal_2008} list 89 members identified in their SACY survey.  Newly proposed members push the total number to more than 100 stars \citep{Schlieder+Lepine+Simon_2010, Schlieder+Lepine+Simon_2012a, Zuckerman+Rhee+Song+etal_2011, Shkolnik+Anglada+Liu+etal_2012, Bowler+Liu+Shkolnik+etal_2012}.  The AB Dor group also covers the entire celestial sphere, with many proposed members in the north.

The age of AB Dor has been revisited and revised many times in the literature.  Ages between 50 and 150 Myr have been derived using HR diagram studies, lithium depletion, activity, and detailed observations of the AB Doradus quadruple system \citep{Zuckerman+Song+Bessel_2004,Torres+Quast+Melo+etal_2008,Mentuch+Brandeker+vanKerwijk+etal_2008,Janson+Brandner+Lenzen+etal_2007,Close+Thatte+Nielsen+etal_2007}.  Several studies argue for a common origin of the AB Dor group and Pleiades open cluster \citep{Luhman+Stauffer+Mamajek_2005, Ortega+Jilinski+delaReza+etal_2007}.

\cite{Barenfeld+Bubar+Mamjek+etal_2013} performed a chemical and kinematic analysis of proposed members and found strong evidence for a kinematic nucleus and associated stream. They caution, however, that their traceback studies and observed chemical inhomogeneity of the proposed members suggest a significant fraction of impostors.  \citeauthor{Barenfeld+Bubar+Mamjek+etal_2013} also place a lower limit of 110 Myr on the group's age by using pre-main sequence contraction times of reliable K-type members.  We combine this well constrained age limit with the previous results showing similarities to the Pleiades to adopt the Pleiades age of 130$\pm$20 Myr \citep{BarradoyNavascues+Stauffer+Jayawardhana+etal_2004} for the AB Doradus moving group.

\subsection{The $\beta$ Pictoris Moving Group}
\label{subsec:BetaPic}

\cite{BarradoyNavascues+Stauffer+Song+etal_1999} identified two young M dwarfs having proper motions consistent with the prototypical debris disk, and now-known planet host, $\beta$ Pictoris \citep{Lagrange+Gratadour+Chauvin+etal_2009, Lagrange+Bonnefoy+Chauvin+etal_2010}; they estimated a system age of $20 \pm 10$ Myr via comparisons to theoretical isochrones.  This led to a search for more stars with similar age and kinematics near $\beta$ Pic by \cite{Zuckerman+Song+Bessell+etal_2001}. They identified 18 systems and coined the name the $\beta$ Pictoris moving group.  \cite{Torres+Quast+daSilva+etal_2006, Torres+Quast+Melo+etal_2008} proposed many $\beta$ Pic members in their SACY survey, while other searches have since proposed the first isolated brown-dwarf member and several additional low-mass members \citep{Lepine+Simon_2009, Rice+Faherty+Cruz_2010, Schlieder+Lepine+Simon_2010, Schlieder+Lepine+Simon_2012a,Schlieder+Lepine+Simon_2012b,Kiss+Moor+Szalai+etal_2011,Malo+Doyon+Lafreniere+etal_2013}.

\cite{Torres+Quast+Melo+etal_2008} list 48 high probability members of $\beta$ Pic; newer additions bring the total to more than 60 stars.  $\beta$ Pic members are spread over the sky with the majority at southern declinations.  The galactic kinematics and age of the group are similar to those of the TW Hydrae association (see \S~\ref{subsec:TWHydrae}), and both groups may be related to star formation in Sco-Cen OB association subgroups \citep{Mamajek+Feigelson_2001,Ortega+delaReza+Jilinski+etal_2002}.

The age of the $\beta$ Pic group has been estimated at 10--20 Myr from HR diagrams, comparison to evolution models, lithium depletion, and kinematics \citep{BarradoyNavascues+Stauffer+Song+etal_1999,Zuckerman+Song+Bessell+etal_2001,Ortega+delaReza+Jilinski+etal_2002,Mentuch+Brandeker+vanKerwijk+etal_2008}.  Two more recent evaluations of the group age include a study of the lithium depletion boundary by \cite{Binks+Jeffries_2013} and a reanalysis of the kinematic age in \cite{Soderblom+Hillenbrand+Jeffries+etal_2013}. \citeauthor{Binks+Jeffries_2013} constrain the age to $21 \pm 4$ Myr  by comparing the minimum luminosity (i.e.~minimum mass) of M-dwarf members that have fully burned their primordial lithium to predictions from evolutionary models. \citeauthor{Soderblom+Hillenbrand+Jeffries+etal_2013} provide a new analysis of proposed member kinematics using revised Hipparcos astrometry and find that the group was not appreciably smaller any time in the past, excluding traceback as a useful dating method in this case.  A detailed analysis by \cite{Jenkins+Pavlenko+Ivanyuk+etal_2012} also provides an age of $\sim$20 Myr for the substellar host \citep{Biller+Liu+Wahhaj+etal_2010} and $\beta$ Pic member, PZ Tel.  We thus adopt the lithium depletion boundary age of $21 \pm 4$ Myr for our analyses.

\subsection{The Castor Moving Group}
\label{subsec:Castor}

The Castor moving group was originally proposed by \cite{Anosova+Orlov_1991} in their study of the dynamical evolution of several multiple systems in the solar neighborhood.  They searched the Catalog of Nearby Stars \citep{Gliese_1969} for all systems inside a velocity cube 6 km\,s$^{-1}$ on a side, centered on the Castor sextuple system.  They found 13 additional stars in 9 systems, and proposed that these stars, together with the Castor system, constitute a moving group.

\cite{BarradoyNavascues_1998} revisited the proposed members of the Castor moving group and performed a more rigorous analysis using new kinematic measurements and age indicators.  They began with a sample of 26 candidate members and found that only 16 met their kinematic and age criteria, which were based on isochrones, activity, and lithium depletion. \citeauthor{BarradoyNavascues_1998} assigned an age of $200 \pm 100$ Myr to the group using the age of proposed member Fomalhaut and its companion TW PsA.  The work of \cite{Montes+Lopez-Santiago+Galvez+etal_2001} led to the identification of eight possible late-type members of Castor, while \cite{Caballero_2010} and \cite{Shkolnik+Anglada+Liu+etal_2012} present additional candidates.

The ages of several original Castor members have been recently reassessed using modern techniques.  \cite{Yoon+Peterson+Kurucz+etal_2010} redetermined the age of Vega to be $455 \pm 13$ Myr using spectroscopic, photometric, and interferometric data together with isochrones.  A full interferometric analysis by Monnier et al.~(2012) increased this age to $\sim$700 Myr. \cite{Mamajek_2012} revisited the age of Fomalhaut and its wide stellar companion and used modern isochrones, lithium depletion measurements, and age/rotation/activity diagnostics to assign them an age of $440 \pm 40$ Myr.  These new results are incompatible with the proposed age of the Castor moving group, and cast doubt on its physical reality as a coeval association. 

In phase space, the Castor moving group lacks a discernible core or tight nucleus of members (velocity dispersion $\sim$1 km\,s$^{-1}$).  Although this may be due to its older age, it may also indicate that the Castor moving group is really a complex of kinematically similar stars with a spread of ages.  \cite{Zuckerman+Vican+Song+etal_2013} and \cite{Mamajek+Bartlett+Seifahrt+etal_2013} reach the latter conclusion and reject a common age for Castor.  Thus, we do not assign the proposed group age to the candidate members in the SEEDS sample, relying instead on single-star age indicators such as activity and lithium depletion.  

\subsection{The Columba Association}
\label{subsec:Columba}

\cite{Torres+Quast+Melo+etal_2008} discovered the Columba association in their SACY survey.  Its kinematics and age are very similar to the Tucana-Horologium association (see \S~\ref{subsec:TucHor}), but it is considered to be kinematically distinct due to its significantly different $W$ velocity.  \citeauthor{Torres+Quast+Melo+etal_2008} proposed 53 members of this association, including some stars originally proposed as members of Tucana-Horologium.  

An additional 14 Columba members were proposed by \cite{Zuckerman+Rhee+Song+etal_2011}.  Their list included many high-mass stars including HR 8799 and $\kappa$ Andromedae, two stars hosting substellar companions \citep{Marois+Macintosh+Barman+etal_2008, Marois+Zuckerman+Konopacky+etal_2010, Carson+Thalmann+Janson+etal_2013}.  \cite{Malo+Doyon+Lafreniere+etal_2013} performed a Bayesian analysis on the full sample of proposed candidates, finding 21 high-probability members on the basis of complete kinematic data.  

The Columba association received some scrutiny in a kinematic study, which questioned HR 8799's membership due to its distance from the bulk of the association throughout an epicyclic orbit simulation \citep{Hinz+Rodigas+Kenworthy+etal_2010}.  \citeauthor{Hinz+Rodigas+Kenworthy+etal_2010} also suggest that since the proposed members of Columba cover such a large volume  of space \citep[$>$100 pc,][]{Torres+Quast+Melo+etal_2008}, it is more likely to be a complex of young stars with a range of ages.  \citeauthor{Torres+Quast+Melo+etal_2008} also noted the Columba association's large spatial extent, as a result of which membership probabilities for this group were significantly lower than for the more compact Tucana-Horologium association.

While the physical reality of the association may not yet be well-established, the stars proposed as members are still excellent targets for direct imaging due to their relative proximity and young ages.  We carefully investigate the age of each target member to verify that it is comparable to the $30^{+20}_{-10}$ Myr age \citep{Marois+Zuckerman+Konopacky+etal_2010} of the group.   

\subsection{The Hercules-Lyra Association}
\label{subsec:HercLyra}

The first indication of this young kinematic group was found by \cite{Gaidos_1998} in their study of young solar analogs.  \citeauthor{Gaidos_1998} identified 5 nearly comoving young stars with a radiant in the constellation Hercules, calling them the Hercules Association.  

\cite{Fuhrmann_2004} studied nearby stars of the galactic disk and halo to identify more stars with kinematics and ages similar to those identified by \citeauthor{Gaidos_1998}.  The resulting updated sample of 15 stars had a radiant point at the border between the constellations Hercules and Lyra, and the Hercules association was renamed the Hercules-Lyra association.  The stars in the \cite{Fuhrmann_2004} sample exhibit rotations, activities, and lithium depletions that suggest generally young ages.  Some stars appeared to be coeval with proposed Ursa Majoris moving group members ($\sim$200 Myr, at the time), while others appeared younger or older, suggesting that the Hercules-Lyra association may not be coeval.

\cite{Lopez-Santiago+Montes+Crespo-Chacon+etal_2006} revisited the proposed Hercules-Lyra association, searching their list of late-type members of kinematic groups \citep{Montes+Lopez-Santiago+Galvez+etal_2001} for new candidates.  They required Galactic UV velocities within 6 km\,s$^{-1}$ of the mean values from \cite{Fuhrmann_2004} but imposed no restriction on the W component of the velocity.  From their initial sample of 27 candidates, \citeauthor{Lopez-Santiago+Montes+Crespo-Chacon+etal_2006} found only 10 meeting their kinematic, lithium, and photometric criteria.  They assigned an age of 150-300 Myr to the association due to consistent results from both lithium abundances and color-magnitude diagrams.  \cite{Shkolnik+Anglada+Liu+etal_2012} proposed an additional low-mass candidate. 

\cite{Eisenbeiss+Ammler-vonEiff+Roell+etal_2013} revisit the membership, age, and multiplicity of the previously proposed members and find only seven systems that meet all of their membership criteria. These stars exhibit Galactic velocity dispersions $>$3.5 km s$^{-1}$ and have ages of $\sim$260$\pm$50 Myr estimated from gyrochronology.  As for the Castor moving group (see Section \ref{subsec:Castor}), the small number and large velocity dispersion of reliably proposed members cast doubt on Hercules-Lyra as a true young stellar association.  We therefore rely on youth indicators such as lithium and chromospheric activity to derive ages for each individual star.

\subsection{The IC 2391 Supercluster}
\label{subsec:IC2391}

\cite{Eggen_1991} noticed that more than 60 field stars and members of the IC 2391 open cluster all have motions directed toward a single convergent point.  Color-magnitude diagrams and comparisons to available isochrones suggested a bimodal age distribution, with one subgroup at $\sim$80 Myr and the other at $\sim$250 Myr.  Further comments on this kinematic group can be found in \cite{Eggen_1992, Eggen_1995}.  

\cite{Montes+Lopez-Santiago+Galvez+etal_2001} reassessed previously proposed members of the IC 2391 supercluster and searched for new late-type candidates using updated astrometry, photometry, and spectroscopy.  After adopting a cluster age of 35--55 Myr from \cite{Eggen_1995}, only 15 stars met their kinematic criteria.  \cite{Maldonado+Martinez-Arnaiz+Eiroa+etal_2010} used similar techniques to search for new members of several proposed kinematic groups, including the IC 2391 supercluster.  In addition to compiling literature data, they performed follow-up spectroscopy to measure radial velocities and stellar age indicators.  They found that when strict kinematic and age criteria were employed, only 5 of 19 candidates remained as probable members.  Furthermore, they caution that the supercluster may have two subgroups mixed in the $UV$ velocity plane, one with an age of $\sim$200--300 Myr, and an older, $\sim$700 Myr component \citep{Lopez-Santiago+Montes+Galvez-Ortiz+etal_2010}.

Unfortunately, much of the existing literature disputes the physical reality of a coeval IC 2391 supercluster.  Strict kinematic and age requirements give a sample with as few as five members, while the proposed ages for members vary by up to a factor of $\sim$20.  We therefore consider claimed IC 2391 supercluster membership as a poor determinant of age and defer to each individual star's age indicators.

\subsection{The Local Association}
\label{subsec:LocalAssociation}

\citeauthor{Eggen_1961} first noticed that several open clusters had galactic kinematics similar to the Pleiades (The Pleiades group).  Eggen later identified more stars with similar kinematics, and proposed the Local Association \citep{Eggen_1975, Eggen_1983b, Eggen_1983c}. 
This kinematic stream included classical clusters such as the Pleiades, $\alpha$ Persei, and Scorpius-Centaurus, along with more than 100 other stars in a large volume of space around the Sun.  The age of the stream was not well-defined, and subsequently spanned the estimated age ranges of its constituent clusters ($\sim$20 to $\sim$150 Myr).
     
\cite{Jeffries+Jewell_1993} studied the kinematics of X-ray and EUV selected late-type stars within 25 pc to identify more than 10 candidate members of the Local Association.  A follow-up survey measured lithium abundances and rotational velocities \citep{Jeffries_1995}.  Seventeen of their late-type candidates had age indicators and kinematics consistent with the Local Association.  \cite{Montes+Lopez-Santiago+Fernandez-Figueroa+etal_2001, Montes+Lopez-Santiago+Galvez+etal_2001} used similar techniques to search for new members, identifying seven stars with spectroscopic youth indicators out of 45 previously proposed candidates.

Although the proposed members of the Local Association do have similar galactic motions, the dispersion in UVW velocities is quite large ($\sim$20 km\,s$^{-1}$), the ages of constituent stars vary by $\sim$100 Myr, and the members are spread out over $\sim$150 pc.  These features disfavor a common origin of the association, and in fact, many of the younger (and much better-defined) moving groups have ages and kinematics placing them within the bounds of the Local Association.  We therefore do not use Local Association membership to infer a star's age, relying instead on individual members' other age indicators.

\subsection{The Tucana-Horologium Association}
\label{subsec:TucHor}

\cite{Zuckerman+Webb_2000} searched the {\it Hipparcos} catalog in the neighborhoods of a few dozen stars with 60 $\mu$m {\it IRAS} excesses, selecting targets with distances and proper motions similar to those of the infrared sample.  Follow-up spectroscopy of these candidates led to the discovery of the Tucanae association, a well-defined kinematic group of stars $\sim$45 pc from the Sun with an age of about 40 Myr.  Nearly simultaneously, \cite{Torres+daSilva+Quast+etal_2000} searched for kinematically similar, X-ray bright stars near the active star EP Eri.  Spectroscopic follow-up of active candidates revealed about 10 stars with very similar kinematics and spectroscopic youth indicators.  These stars, comprising the Horologium association, had an isochronal age of $\sim$30 Myr and distances of $\sim$60 pc.  Since the Tucanae and Horologium associations have similar kinematics and the same estimated age, they were later merged to form the Tucana-Horologium association \citep{Zuckerman+Song+Webb_2001}. 

\cite{Zuckerman+Song_2004} listed 31 proposed members of Tucana-Horologium.  \cite{Torres+Quast+Melo+etal_2008} identified 13 additional members in their SACY survey, bringing the total to 44.  In the same review, \citeauthor{Torres+Quast+Melo+etal_2008} associated Tucana-Horologium with two more recently discovered associations of similar age---Columba (see \S~\ref{subsec:Columba} and Carina---and suggested that these three groups together form a large complex of young stars (the Great Austral Young Association, or GAYA).  \cite{Zuckerman+Rhee+Song+etal_2011} proposed several new members, including the first at northern declinations.  \cite{Malo+Doyon+Lafreniere+etal_2013} have also presented a list of high-probability, low-mass candidate members.  The value of these new candidates is exemplified by the recent imaging discovery of a very novel triple system comprised of two late M dwarf Tucana-Horologium candidates and a 12-14 $M_J$ substellar companion \citep{Delorme+Gagne+Girard+etal_2013}. 

The Tucana-Horologium association is one of the best-studied nearby young groups.  Most of its proposed members are spatially and kinematically well-defined with little scatter in velocity space.  An age of $\sim$30 Myr is consistently derived for its members; we adopt $30^{+10}_{-20}$ Myr \citep{Zuckerman+Song+Webb_2001} as the age of the group.

\subsection{The TW Hydrae Association} 
\label{subsec:TWHydrae}

The TW Hydrae association, proposed by \cite{Kastner+Zuckerman+Weintraub+etal_1997}, was the first very young moving group to be discovered.  Early work by \cite{Rucinski+Krautter_1983} demonstrated that the nearby star TW Hya exhibited classical T-Tauri properties.  The release of the {\it IRAS} point source catalog \citep{Helou_Walker_1988} led to spectroscopic surveys of field stars with mid-IR excesses \citep{delaReza+Torres+Quast+etal_1989, Gregorio-Hetem+Lepine+Quast+etal_1992}.  These surveys identified four additional T-Tauri stars near TW Hya, and suggested that they may be members of a nearby T-Tauri association.  \cite{Kastner+Zuckerman+Weintraub+etal_1997} later confirmed the five stars' common age by their strong X-ray emission and lithium absorption.

\cite{Webb+Zuckerman+Platais+etal_1999} surveyed X-ray bright targets near TW Hydrae to identify additional members of the group.  Subsequent surveys and analyses have since brought the number of proposed members to about 30 \citep{Zuckerman+Webb+Schwartz+etal_2001, Gizis_2002, Reid_2003, Torres+Quast+deLaReza+etal_2003, Zuckerman+Song_2004, Mamajek_2005, BarradoyNavascues_2006, Torres+Quast+Melo+etal_2008}.  One notable member is 2M1207, a young brown dwarf with a directly imaged planetary mass companion \citep{Chauvin+Lagrange+Dumas+etal_2004}. The age of the association has been determined using many different methods, including HR diagram placement, H$\alpha$ diagnostics, lithium depletion, and kinematics; the most commonly cited age is $\sim$8 Myr.

More recent work on the TW Hydrae association has focused on identifying new, low-mass members. \cite{Looper+Burgasser+Kirkpatrick+etal_2007, Looper+Bochanski+Burgasser+etal_2010, Looper+Mohanty+Bochanski+etal_2010} identified three late M type members, two of which host accretion disks.  \cite{Rodriguez+Bessell+Zuckerman+etal_2011} and \cite{Shkolnik+Liu+Reid_2011} used UV excesses as observed by the {\it GALEX} satellite to select low-mass candidate members, while \cite{Schneider+Song+Melis+etal_2012} used IR excesses measured by the {\it WISE} satellite.  Parallaxes for many proposed members were measured by \cite{Weinberger+Anglada+Escude+etal_2013}, who found that the association resembles an extended filament with an average member distance of 56 pc.  These distance measurements enable precise HR diagram placement and comparison to model isochrones. A Gaussian fit to the isochrone-based age distribution provides a mean age of 9.5$\pm$5.7 Myr.

Despite the extensive study of the classical young association, TW Hydrae's evolution and membership are still being refined.  Searches for new members continue (e.g., \citealt{Malo+Doyon+Lafreniere+etal_2013}), and may eventually lead to a complete census of this youngest and closest association. For our analyses, we adopt an age of 10$\pm$5 Myr for the group.

\subsection{The Ursa Major or Sirius Supercluster}
\label{subsec:UrsaMajor}

The literature is rich with references to a kinematic association of stars related to the constellation Ursa Major, first introduced in the 19$^{\rm th}$ century by \cite{Proctor_1869}.  A complete history of these stars is beyond the scope of this paper; however, we do mention prominent studies and refer the reader to references found therein for a complete review. We aim to establish in this subsection a distinction between the coeval Ursa Majoris moving group and a dynamical stream of stars with generally consistent kinematics but heterogeneous ages known as the Ursa Major or Sirius supercluster.

The most modern and comprehensive study of the Ursa Majoris moving group is \cite{King+Villarreal+Soderblom+etal_2003}, which reevaluated previously proposed members using new astrometry, photometry, and spectroscopy.  From an input list of $\sim$220 proposed Ursa Majoris candidates compiled from various sources, \citeauthor{King+Villarreal+Soderblom+etal_2003} identified 57 probable and possible members that are well defined in kinematic and color-magnitude space.  Comparison of evolution models to the color-magnitude diagram of their refined membership list suggests an age of $500 \pm 100$ Myr for the group.  \cite{Shkolnik+Anglada+Liu+etal_2012} later identified four additional candidate M dwarf members.  Since the Ursa Majoris moving group contains a well defined nucleus with small velocity dispersions and is well characterized in a color-magnitude diagram, the estimated age of the group can be reliably applied to stars that meet membership criteria.

The Sirius supercluster was originally proposed as a remnant kinematic stream associated with the Ursa Majoris moving group nucleus by \cite{Eggen_1958}. Further members were proposed by \cite{Palous+Hauck_1986}, who estimated an isochronal age of $\sim$490 Myr and proposed that the stars are chemically homogeneous. \cite{Famaey+Jorissen+Luri+etal_2005, Famaey+Siebert+Jorissen_2008} present modern analyses of the proposed Sirius supercluster and other superlcusters associated to well defined, coeval associations (Hyades, Pleiades) using new Hipparcos and Tycho-2 astrometry and radial velocity data from the CORAVEL spectrometer. Their analyses find that kinematically consistent members of the proposed superclusters do not have consistent isochronal ages. They propose that these structures in kinematic space are stellar streams likely generated by dynamical perturbations and are comprised of stars with heterogeneous ages that were not products of the same star formation event. Thus, kinematic membership to the Ursa Major or Sirius supercluster, in contrast to the well defined Ursa Majoris moving group, is not useful as a stellar age indicator. We therefore do not assign the proposed supercluster age to the possible member we observed (HIP 73996) but rather rely on an individually assigned age from our own and literature measurements.

\subsection{Target List and Selection Criteria}
\label{subsec:TargetSelection}

\begin{figure}
\centering\includegraphics[width=0.9\linewidth]{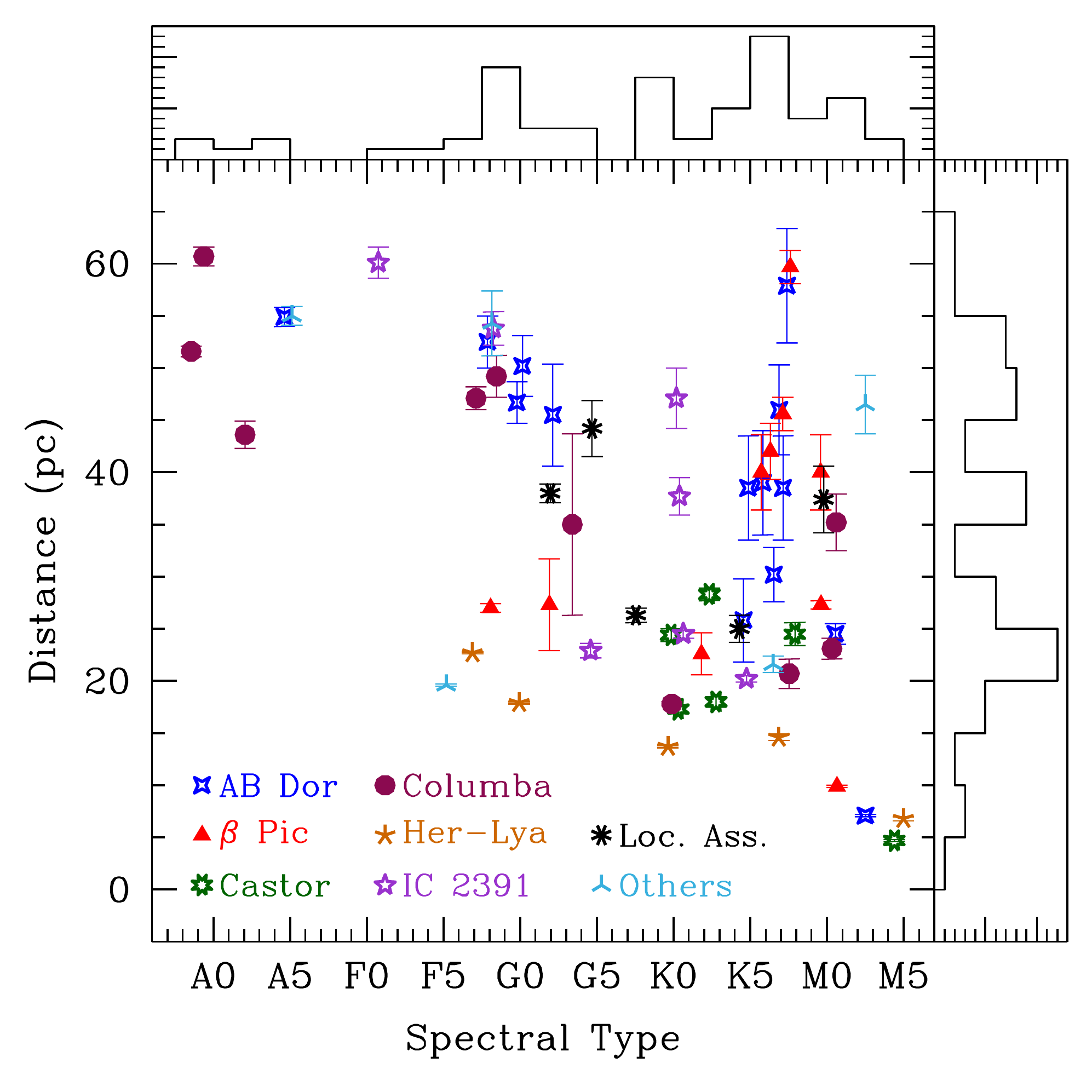}
\caption{Distances, spectral types, and host moving group for our target sample.  One star, HIP 78557 (spectral type G0), has a trigonometric distance of $82 \pm 10$ pc, placing it outside of the plot.  Fifty-one of our 63 targets are within 50 pc, and all but three are within 60 pc, while their spectral types range from late B to early M.}
\label{fig:target_summary}
\end{figure}

Table \ref{tab:stellarprops} lists the SEEDS Moving Groups targets in order of right ascension.  Figure \ref{fig:target_summary} shows the targets' distances and spectral types.  Fifty-one out of 63 targets are within 50 pc, and all but three are within 60 pc.  The spectral types of the main MG sample vary from late F to early M, equivalent to a range of roughly $0.4$--$1.3$ M$_\odot$.  We also list five stars, HIP 23362, HIP 32104, HIP 83494, HIP 93580, and HIP 116805 (= $\kappa$ And), which are more massive A and early B stars selected for the high-mass star sample, but which have been suggested to belong to young moving groups.  

The main SEEDS MG targets were selected according to the following criteria, in order of priority:
\begin{enumerate}
\item Identification with a young moving group ($\lesssim$500 Myr, with younger targets preferred)
\item Proximity to Earth
\item Mass ($\sim$1 $M_\odot$ preferred)
\item Lack of a close binary companion
\item Lack of archival high-contrast observations
\item $R$-magnitude $< 12$ (for AO performance)
\item Declination $> -25^\circ$
\item Field rotation in one hour of observing time
\item $H$-magnitude $\gtrsim 5$ (to limit saturation)
\item High Galactic latitude (to limit chance alignments)
\end{enumerate}

Targets were proposed before each observing run and observed as permitted by conditions and priorities for other SEEDS categories.

\begin{deluxetable*}{lcccccccccr}
\tablewidth{0pt}
\vspace{-0.6truein}
\tablecaption{The SEEDS Moving Group Target List: Basic Stellar Properties}
\tablehead{
    \multicolumn{4}{c}{Designations} &
    \colhead{$\alpha$ (J2000)\tablenotemark{a}} &
    \colhead{$\delta$ (J2000)\tablenotemark{a}} &
    \colhead{Distance\tablenotemark{a}} &
    \colhead{Spectral} &
    \colhead{$V$\tablenotemark{b}} &
    \colhead{$H$\tablenotemark{c}} & 
    \colhead{Moving} \\
    \colhead{HIP$\!\!\!\!\!\!\!\!\!\!\!\!\!\!\!\!\!\!\!\!\!\!$} &
    \colhead{HD} &
    \colhead{GJ} &
    \colhead{Other} &
    \colhead{(h m s)} &
    \colhead{($^\circ$ $'$ $''$)} &
    \colhead{(pc)} &
    \colhead{Type\tablenotemark{f}} &
    \colhead{(mag)} &
    \colhead{(mag)} &
    \colhead{Group}
}
    
\startdata
   544 &    166 &      5 &        V439 And & 00 06 36.8 & $+$29 01 17 & $13.7 \pm 0.1$ &    G8V (1) &  6.06 & 4.63 & Her Lya      \\    
  1134 &    984 & \ldots &          \ldots & 00 14 10.3 & $-$07 11 57 & $47.1 \pm 1.1$ &    F7V (2) &  7.32 & 6.17 & Columba      \\    
\ldots & \ldots & \ldots &          FK Psc & 00 23 34.7 & $+$20 14 29 & $59.7 \pm 1.6$\tablenotemark{k} &  K7.5V (3) & 10.84 & 7.50 & $\beta$ Pic \\    
  3589 &   4277 & \ldots &       BD+54 144 & 00 45 50.9 & $+$54 58 40 & $52.5 \pm 2.5$ &    F8V\tablenotemark{e} &  7.81 & 6.40 & AB Dor       \\    
  4979 &  6288A & \ldots &          26 Cet & 01 03 49.0 & $+$01 22 01 & $60.1 \pm 1.5$ &    A8IV (2) &  6.07 & 5.51 & IC 2391      \\    
  6869 &   8941 & \ldots &          \ldots & 01 28 24.4 & $+$17 04 45 & $53.8 \pm 1.6$ & F8IV-V (4) &  6.60 & 5.40 & IC 2391      \\    
\ldots & \ldots & \ldots &          HS Psc & 01 37 23.2 & $+$26 57 12 &         $38.5$\tablenotemark{k} &     K5Ve (5) & 10.72 & 7.78 & AB Dor       \\    
 10679 & 14082B & \ldots &      BD+28 382B & 02 17 24.7 & $+$28 44 30 & $27.3 \pm 4.4$ &    G2V\tablenotemark{e} &  7.76 & 6.36 & $\beta$ Pic  \\    
\ldots & \ldots & \ldots & BD+30 397B & 02 27 28.0 & $+$30 58 41 & $40.0 \pm 3.6$ &     M0 (6) & 12.44 & 8.14 & $\beta$ Pic  \\    
 11437 & \ldots & \ldots &          AG Tri & 02 27 29.3 & $+$30 58 25 & $40.0 \pm 3.6$ &    K7V (7) & 10.08 & 7.24 & $\beta$ Pic  \\    
 12545 & \ldots & \ldots &       BD+05 378 & 02 41 25.9 & $+$05 59 18 & $42.0 \pm 2.7$ &    K6Ve (8) & 10.20 & 7.23 & $\beta$ Pic  \\    
 12638 &  16760 & \ldots &          \ldots & 02 42 21.3 & $+$38 37 07 & $45.5 \pm 4.9$ &     G2\tablenotemark{e} &  8.77 & 7.10 & AB Dor       \\    
 12925 &  17250 & \ldots &       BD+04 439 & 02 46 14.6 & $+$05 35 33 & $54.3 \pm 3.1$ &     F8\tablenotemark{e} &  7.88 & 6.63 & Tuc-Hor      \\    
 17248 & \ldots & \ldots &          \ldots & 03 41 37.3 & $+$55 13 07 & $35.2 \pm 2.7$ &     M0.5 (9) & 11.20 & 7.65 & Columba      \\    
 23362 &  32309 & \ldots &         HR 1621 & 05 01 25.6 & $-$20 03 07 & $60.7 \pm 0.9$ &    B9V (10) &  4.88 & 5.02 & Columba  \\ 
 25486 &  35850 & \ldots &          AF Lep & 05 27 04.8 & $-$11 54 03 & $27.0 \pm 0.4$ &    F8V (11) &  6.30 & 5.09 & $\beta$ Pic  \\    
\ldots &  36869 & \ldots &          AH Lep & 05 34 09.2 & $-$15 17 03 & $35.0 \pm 8.7$\tablenotemark{*} &    G2V (10) &  8.45 & 6.98 & Columba   \\    
 29067 & \ldots &   9198 &          \ldots & 06 07 55.3 & $+$67 58 37 & $24.5 \pm 1.1$ &    K6V (1) &  9.74 & 6.81 & Castor       \\    
 30030 &  43989 & \ldots &       V1358 Ori & 06 19 08.1 & $-$03 26 20 & $49.2 \pm 2.0$ &    G0V (2) &  7.95 & 6.59 & Columba      \\    
 32104 &  48097 & \ldots &          26 Gem & 06 42 24.3 & $+$17 38 43 & $43.6 \pm 1.3$ &    A2V (12) &  5.22 & 5.07 & Columba  \\
\ldots & \ldots & \ldots &        V429 Gem & 07 23 43.6 & $+$20 24 59 & $25.8 \pm 4.0$\tablenotemark{s} &   K5V (13) &  10.03 & 7.03 & AB Dor       \\    
 37288 & \ldots &    281 &          \ldots & 07 39 23.0 & $+$02 11 01 & $14.6 \pm 0.3$ &     K7 (14) &  9.66 & 6.09 & Her Lya      \\    
 39896 & \ldots &  1108A &          FP Cnc & 08 08 56.4 & $+$32 49 11 & $20.7 \pm 1.4$ &     K7 (14) &  9.99 & 6.58 & Columba    \\    
 40774 & \ldots & \ldots &        V397 Hya & 08 19 19.1 & $+$01 20 20 & $22.9 \pm 0.7$ &    G5V\tablenotemark{e} &  8.35 & 6.22 & IC 2391      \\    
 44526 &  77825 & \ldots &        V405 Hya & 09 04 20.7 & $-$15 54 51 & $28.3 \pm 0.6$ &    K3V (11) &  8.78 & 6.54 & Castor       \\    
 45383 &  79555 &    339 &          \ldots & 09 14 53.7 & $+$04 26 34 & $18.0 \pm 0.5$ &    K3V (1) &  7.96 & 5.40 & Castor       \\    
 46843 &  82443 &  354.1 &          DX Leo & 09 32 43.8 & $+$26 59 19 & $17.8 \pm 0.2$ &    K1V (11) &  7.06 & 5.24 & Columba      \\    
 50156 & \ldots &   2079 &          DK Leo & 10 14 19.2 & $+$21 04 30 & $23.1 \pm 1.0$ &  M0.7V (15) & 10.13 & 6.45 & Columba      \\    
\ldots & \ldots &    388 &          AD Leo & 10 19 36.3 & $+$19 52 12 &  $4.7 \pm 0.1$ &  M3 (14) &  9.46 & 4.84 & Castor       \\    
 50660 & \ldots & \ldots &      NLTT 24062 & 10 20 45.9 & $+$32 23 54 & $47.1 \pm 2.9$ &    K0V\tablenotemark{e} &  9.18 & 7.38 & IC 2391      \\    
 51317 & \ldots &    393 &        LHS 2272 & 10 28 55.6 & $+$00 50 28 &  $7.1 \pm 0.1$ &  M2.5V (16) &  9.59 & 5.61 & AB Dor \\    
 53020 & \ldots &    402 &          EE Leo & 10 50 52.0 & $+$06 48 29 &  $6.8 \pm 0.2$ &  M5.0V (16) & 11.68 & 6.71 & Her Lya      \\    
 53486 &  94765 &   3633 &          GY Leo & 10 56 30.8 & $+$07 23 19 & $17.3 \pm 0.3$ &    K2.5V (1) &  7.37 & 5.35 & Castor       \\    
\ldots &  95174 & \ldots &          \ldots & 10 59 38.3 & $+$25 26 15 & $22.6 \pm 2.0$\tablenotemark{k} &     K2 (17) &  8.46 & 5.96 & $\beta$ Pic  \\    
 54155 &  96064 & \ldots &          HH Leo & 11 04 41.5 & $-$04 13 16 & $26.3 \pm 0.7$ &    G8V (1) &  7.60 & 5.90 & Loc.~Ass.    \\    
\ldots & \ldots & \ldots &           TWA 2 & 11 09 13.8 & $-$30 01 40 & $46.5 \pm 2.8$\tablenotemark{p} &    M2Ve (8) & 11.12 & 6.93 & TW Hya       \\    
\ldots & \ldots & \ldots &  TYC 3825-716-1 & 11 20 50.5 & $+$54 10 09 & $57.9 \pm 5.5$\tablenotemark{k} &     K7 (18) & 12.14 & 8.69 & AB Dor       \\    
 59280 & 105631 &   3706 &         G 123-7 & 12 09 37.3 & $+$40 15 07 & $24.5 \pm 0.4$ &    G9V (1) &  7.46 & 5.70 & IC 2391      \\    
\ldots & \ldots & \ldots &  TYC 4943-192-1 & 12 15 18.4 & $-$02 37 28 & $30.2 \pm 2.6$\tablenotemark{s} &     M0Ve (5) & 11.34 & 8.00 & AB Dor       \\    
 60661 & \ldots &    466 &          \ldots & 12 25 58.6 & $+$08 03 44 & $37.4 \pm 3.2$ &    M0V (19) &  10.29 & 7.31 & Loc.~Ass.    \\    
 63317 & 112733 & \ldots &          \ldots & 12 58 32.0 & $+$38 16 44 & $44.2 \pm 2.7$ &    K0V (19) &  8.64 & 6.95 & Loc.~Ass.       \\    
\ldots & \ldots & \ldots &          FH CVn & 13 27 12.1 & $+$45 58 26 & $46.0 \pm 4.3$\tablenotemark{k} &     K7 (18) & 11.16 & 8.20 & AB Dor       \\    
 66252 & 118100 &    517 &          EQ Vir & 13 34 43.2 & $-$08 20 31 & $20.2 \pm 0.3$ &    K4.5V (1) &  9.25 & 6.31 & IC 2391      \\    
 67412 & 120352 & \ldots &          \ldots & 13 48 58.2 & $-$01 35 35 & $37.7 \pm 1.8$ &    G8V (2) &  8.51 & 6.89 & IC 2391      \\    
 73996 & 134083 &    578 &           c Boo & 15 07 18.1 & $+$24 52 09 & $19.6 \pm 0.1$ &    F5V (1) &  4.93 & 4.01 & UMa          \\    
 78557 & 143809 & \ldots &      BD+04 3100 & 16 02 22.4 & $+$03 39 07 &    $82 \pm 10$ &     G0V (2) &  8.77 & 7.52 & Loc.~Ass.       \\    
 82688 & 152555 & \ldots &          \ldots & 16 54 08.1 & $-$04 20 25 & $46.7 \pm 2.0$ & F8/G0V (2) &  7.82 & 6.48 & AB Dor       \\    
 83494 & 154431 & \ldots &         HR 6351 & 17 03 53.6 & $+$34 47 25 & $55.0 \pm 0.9$ &    A5V (12) &  6.08 & 5.68 & Tuc-Hor  \\
 87579 & \ldots &    697 &          \ldots & 17 53 29.9 & $+$21 19 31 & $24.4 \pm 0.6$ &     K2.5V (1) &  8.50 & 6.30 & Castor       \\    
 87768 & \ldots &    698 &          \ldots & 17 55 44.9 & $+$18 30 01 & $25.0 \pm 1.3$ &    K5V (1) &  9.24 & 6.42 & Loc.~Ass.    \\    
 91043 & 171488 & \ldots &        V889 Her & 18 34 20.1 & $+$18 41 24 & $38.0 \pm 0.9$ &    G0V (4) &  7.40 & 5.90 & Loc.~Ass.    \\    
 93580 &  177178 & \ldots &        HR 7214 & 19 03 32.3 & $+$01 49 08 & $54.9 \pm 0.9$ &    A4IV/V (2) &  5.82 & 5.36 & AB Dor  \\
\ldots & \ldots & \ldots &      BD+05 4576 & 20 39 54.6 & $+$06 20 12 &         $38.5$\tablenotemark{k} &     K7Ve (5) & 10.52 & 7.35 & AB Dor       \\    
102409 & 197481 &    803 &          AU Mic & 20 45 09.5 & $-$31 20 27 &  $9.9 \pm 0.1$ &    M1Ve (8) &  8.76 & 4.83 & $\beta$ Pic  \\    
\ldots & 201919 & \ldots &          \ldots & 21 13 05.3 & $-$17 29 13 &           $39$\tablenotemark{k} &    K6Ve (8) & 10.43 & 7.75 & AB Dor       \\    
107350 & 206860 &   9751 &          HN Peg & 21 44 31.3 & $+$14 46 19 & $17.9 \pm 0.1$ &    G0V (11) &  5.95 & 4.60 & Her Lya      \\    
\ldots & \ldots & \ldots & TYC 2211-1309-1 & 22 00 41.6 & $+$27 15 14 & $45.6 \pm 1.6$\tablenotemark{k} &     M0Ve (3) & 11.37 & 7.95 & $\beta$ Pic  \\    
111449 & 213845 &  863.2 &        LTT 9081 & 22 34 41.6 & $-$20 42 30 & $22.7 \pm 0.1$ &    F5V (11) &  5.21 & 4.27 & Her Lya      \\    
114066 & \ldots &   9809 &          \ldots & 23 06 04.8 & $+$63 55 34 & $24.5 \pm 1.0$ &  M0.3V (15) & 10.92 & 7.17 & AB Dor       \\    
115162 & \ldots & \ldots &      BD+41 4749 & 23 19 39.6 & $+$42 15 10 & $50.2 \pm 2.9$ &     G8V (20)\tablenotemark{d} &  8.93 & 7.28 & AB Dor       \\    
\ldots & \ldots & \ldots &    BD$-$13 6424 & 23 32 30.9 & $-$12 15 51 & $27.3 \pm 0.4$\tablenotemark{k} &    M0Ve (8) & 10.69 & 6.77 & $\beta$ Pic  \\      
 116805 &  222439 & \ldots &  $\kappa$ And & 23 40 24.5 & $+$44 20 02 & $51.6 \pm 0.5$ &    B9IVn (21) & 4.13 & 4.60 & Columba  
\enddata

\tablenotetext{a}{Position and parallax from the {\it Hipparcos} catalog \citep{vanLeeuwen_2007} unless otherwise noted}
\tablenotetext{b}{Values taken from the Tycho-2 catalog \citep{Hog+Fabricius+Makarov+etal_2000} and converted to Johnson $V$, with the following exceptions: BD$+$30 397B \citep{Weis_1993}; HIP 53020 \citep{Landolt_1992}.}
\tablenotetext{c}{Values taken from the 2MASS catalog \citep{Cutri+Skrutskie+vanDyk+etal_2003}}
\tablenotetext{d}{Spectral type also discussed in this work}
\tablenotetext{e}{Spectral type listed (but unsourced, or  sourced as SIMBAD) in the {Hipparcos} catalog} 
\tablenotetext{k}{Kinematic distance assuming group membership.  References: FK Psc, TYC 2211-1309-1, BD$-$13 6424 \citep{Lepine+Simon_2009}; HS Psc, BD+05 4576 \citep{Schlieder+Lepine+Simon_2010}; HD 95174, TYC 3825-716-1, FH CVn \citep{Schlieder+Lepine+Simon_2012b}; HD 201919 \citep{Torres+Quast+Melo+etal_2008}}
\tablenotetext{*}{This {\it Tycho} parallax \citep{Hog+Fabricius+Makarov+etal_2000} is far below the distance inferred from spectroscopy \citep[59 pc,~][]{Zuckerman+Rhee+Song+etal_2011}, and may be unreliable.}
\tablenotetext{s}{Spectroscopic parallax.  References: V429 Gem \citep{Reid+Cruz+Allen+etal_2004}; TYC 4943-192-1 \citep{Agueros+Anderson+Covey+etal_2009}}
\tablenotetext{p}{Trigonometric parallax from \cite{Weinberger+Anglada+Escude+etal_2013}}

\tablenotetext{f}{References: 1 \citep{Gray+Corbally+Garrison+etal_2003}; 2 \citep{Houk+Swift_1999}; 3 \citep{Lepine+Simon_2009}; 4 \citep{White+Gabor+Hillenbrand_2007}; 5 \citep{Schlieder+Lepine+Simon_2010}; 6 \citep{Zuckerman+Song_2004}; 7 \citep{Torres+Quast+Melo+etal_2008}; 8 \citep{Torres+Quast+daSilva+etal_2006}; 9 \citep{Zuckerman+Rhee+Song+etal_2011}; 10 \citep{Houk+Smith-Moore_1988}; 11 \citep{Gray+Corbally+Garrison+etal_2006}; 12 \citep{Abt+Morrell_1995}; 13 \citep{Reid+Cruz+Allen+etal_2004}; 14 \citep{Reid+Hawley+Gizis_1995, Hawley+Gizis+Reid_1996}; 15 \citep{Shkolnik+Liu+Reid_2009}; 16 \citep{Jenkins+Ramsey+Jones+etal_2009}; 17 \citep{Scholz+Meusinger+Jahreiss_2005}; 18 \citep{Schlieder+Lepine+Simon_2012b}; 19 \citep{Lopez-Santiago+Montes+Galvez-Ortiz+etal_2010}; 20 \citep{Ofek_2008}; 21 \citep{Wu+Singh+Prugniel+etal_2011}}

\label{tab:stellarprops}
\end{deluxetable*}

\section{Other Age Indicators} 
\label{sec:otherindicators}

The most reliable age dating methods rely on coeval associations of stars, such as kinematic moving groups or globular clusters.  The members of such a coeval association may be placed on a color-magnitude diagram where isochrones of single stellar populations offer extremely reliable age estimates.  Unfortunately, many stars in our sample (and a much larger fraction of other high-contrast imaging surveys) are not reliable members of a coeval association.  For the sample presented here, we consider AB Dor, $\beta$ Pic, Columba, Tuc-Hor, TW Hydrae, and Ursa Majoris as coeval associations (See \S~\ref{sec:moving_groups}).  We rely on the age indicators described below to assign ages to the stars in Castor, Hercules-Lyra, IC 2391, and the Local Association.

All of these single star age indicators rely to some degree on stellar convection and rotation.  Late F-type and later stars have large convective zones, where stellar dynamos generate substantial magnetic fields from differential rotation \citep{Parker_1955, Glatzmaier_1985} and power vigorous chromospheric and coronal activity.  As a star ages, its magnetized wind carries away angular momentum, and the stellar rotation and magnetically-powered activity gradually decrease.  Convection also carries material from the stellar surface down into the hotter interior, where fragile elements and isotopes like $^7$Li are destroyed.

We discuss five individual age indicators in the following sections: chromospheric activity traced by Ca\,{\sc ii} HK emission, coronal activity traced by X-rays, stellar rotation, photospheric lithium abundance, and isochrone fitting.  These indicators have been studied extensively and calibrated using coeval stellar clusters and associations.

\subsection{Chromospheric Activity} 
\label{subsec:chromo}

The stellar chromosphere is a low-density region above the photosphere containing a strong temperature inversion.  Magnetic reconnection is believed to be responsible for heating the chromosphere, which is visible as an emission line spectrum superimposed on the photosphere's continuum and absorption lines \citep{Wilson_1963}.  The chromospheric emission lines are much narrower and fainter than the corresponding photospheric absorption lines.  Two of the stronger lines are Ca II H and K at 3968 \AA and 3934 \AA, with the chromospheric emission line strengths often parameterized by $R'_{\rm HK}$, the ratio of the flux in the emission line cores to that in the underlying photospheric continuum \citep{Noyes+Hartmann+Baliunas+etal_1984}.

Chromospheric activity has long been known to correlate with stellar age on the main sequence; it is dramatically stronger in young clusters than in the Sun and local field stars \citep{Wilson_1963}.  Multi-decade observations \citep{Baliunas+Sokoloff+Soon_1996} have provided activity measurements for hundreds of stars in well-dated young clusters and (presumably coeval) binaries, enabling the calibration of $R'_{\rm HK}$ as an age indicator for young stars.  \cite{Mamajek+Hillenbrand_2008}, hereafter MH08, have recently re-calibrated $R'_{\rm HK}$ as an age indicator.  They find the tightest correlation by first using chromospheric activity to estimate the Rossby number Ro, the ratio of the rotational period to the convective overturn timescale, and then using the Rossby number and $B-V$ color to infer an age.  Practically, this means that the estimated age is a function of both activity and color (i.e.,~mass).  Omitting uncertainties in the fitted parameters and combining Equations (4) and (12)--(14) from MH08, we have
\begin{equation}
\frac{\tau}{\rm Myr} \approx \left( \frac{\tau_C \left[ 0.808 - 2.966 \left(\log R'_{\rm HK} + 4.52 \right) \right]}{0.407 \left(B - V - 0.495 \right)^{0.325}} \right)^{1.767},
\label{eq:rhk_ages}
\end{equation}
where $\tau_C$ is the convective overturn timescale, and is related to $B-V$ color by Equation (4) in \cite{Noyes+Hartmann+Baliunas+etal_1984}:
\begin{equation}
\log \tau_C = 1.362 - 0.166x + 0.025x^2 - 5.323x^3~,
\label{eq:t_conv1}
\end{equation}
with $x \equiv 1 - (B-V)$ and $x > 0$ (spectral type mid K or earlier).  For $x < 0$ (late K and M stars), the fit is
\begin{equation}
\log \tau_C = 1.362 - 0.14x~.
\label{eq:t_conv2}
\end{equation}
Equation \eqref{eq:rhk_ages} applies to ``active'' stars with $-5.0 < \log R'_{\rm HK} < -4.3$.  While nearly every star in the SEEDS Moving Group sample with archival $R'_{\rm HK}$ data satisfies this minimum activity level, many are too active for Equation \eqref{eq:rhk_ages} to provide an accurate age estimate.  Further, this relation requires $B-V \geq 0.5$ (spectral type late F or later), and is poorly calibrated for $B-V \gtrsim 1$.  For some SEEDS targets, chromospheric activity provides only an upper limit on the age, while for others that do not satisfy the color criterion, chromospheric activity is of little value as an age indicator.

MH08 estimate the scatter about Equation \eqref{eq:rhk_ages} using both field binaries and well-dated clusters with ages ranging from 5 Myr to 4 Gyr.  For stars in the ``active'' regime with multi-decade $R'_{\rm HK}$ data, they estimate a scatter of 0.10 in Rossby number Ro, while for single-epoch chromospheric measurements, they estimate a scatter of 0.16.  We use multi-epoch data wherever possible.  While only one of our targets (HIP 107350) has multi-decade Mt.~Wilson data, many have several epochs from \cite{Isaacson+Fischer_2010}.  For targets with more than one single-epoch HK value (but no multi-epoch data), we take the median of the literature values.  In two cases,We expect our precision to be somewhat better than is reflected in a scatter of 0.16 in Ro; however, we provisionally adopt an uncertainty of 0.16 for all but the Mt.~Wilson data.  Very active stars with $R'_{\rm HK} > -4.3$ have much larger uncertainties.  We assign these targets only upper limits on age, using a uniform probability distribution between 0 and the minimum age accessible to chromospheric activity measurements.  

We compile chromospheric activity measurements from a wide variety of sources, using the relations given in \cite{Noyes+Hartmann+Baliunas+etal_1984} to transform all onto the Mt.~Wilson system.  For two stars, HIP 40774 and HIP 50660, the original reference \citep{Strassmeier+Washuett+Granzer+etal_2000} used different units, which were recently calibrated and transformed onto the Mt.~Wilson system \citep{Pace_2013}.  All of our literature $R'_{\rm HK}$ values are listed in Table \ref{tab:stellarages_full}.

\subsection{X-ray Activity} 
\label{subsec:xray_activity}

X-ray activity presents a similar measure of magnetic activity, though this emission comes from the high-temperature stellar corona.  While the coronal heating mechanism remains uncertain and presents formidable modeling challenges \citep{Klimchuk_2006}, it almost certainly involves the deposition of magnetic energy, either from reconnection events \citep[e.g.~][]{Parker_1988, Masuda+Kosugi+Hara+etal_1994} or the dissipation of magento-acoustic and/or Alfv\'en waves \citep[e.g.~][]{Heyvaerts+Priest_1983, Davila_1987}.  As with chromospheric activity, X-ray activity declines as a star ages and loses angular momentum \citep{Hempelmann+Schmitt+Schultz+etal_1995}. 

X-ray activity is typically measured as the ratio of a star's X-ray flux (within the 0.1-2.4 keV {\it ROSAT} bandpass; \citealt{Voges+Aschenbach+Boller+etal_1999}, with a hardness correction) to its bolometric flux.  We use the formula given in \cite{Schmitt+Fleming+Giampapa_1995}:
\begin{equation}
F_{\rm X} = \left( 5.30 {\rm HR} + 8.31 \right) {\rm CR} \times 10^{-12}~{\rm erg~cm^{-2}~s}^{-1}~,
\end{equation}
where the CR is the count rate and HR is the ratio of the difference in count rate between the hard (0.52--2.1 keV) and soft (0.1--0.41 keV) channels to the total count rate.  For targets not detected by {\it ROSAT}, we estimate upper limits on their X-ray fluxes using the exposure time of the nearest detected source (usually $\sim$$0.\!\!^\circ 5$) in the faint source catalog \citep{Voges+Aschenbach+Boller+etal_2000}, requiring no more than 9 expected photons, and assuming a hardness ratio of 0 (roughly the mean of our sample).  A source with 9 expected photons would have a $\sim$90\% probability of producing at least 6 detected photons, the minimum required for inclusion in the {\it ROSAT} catalog.  Combined with a small correction for background subtraction and some uncertainty in the hardness ratio, these $F_{\rm X}$ values should be considered approximate upper limits.  

The indicator $R_{\rm X}$ is $F_{\rm X}$ normalized to a star's bolometric flux.  For G and earlier stars, we convert the $V$ magnitudes in Table \ref{tab:stellarprops} to bolometric fluxes using the relations derived in \cite{Flower_1996}--these were originally misprinted and have been corrected in, e.g., \cite{Torres_2010}.  These bolometric corrections are not valid for M stars; we therefore adopt the bolometric correction of \cite{Kenyon+Hartmann_1995}, which uses $V$, $J$, and $K$ band magnitudes, for K and M dwarfs, adjusting the zero-point of the correction scale accordingly \citep{Torres_2010}.

As for chromospheric activity, MH08 have calibrated an X-ray/color/age relation, equivalent to
\begin{equation}
\frac{\tau}{\rm Myr} \approx \left( \frac{\tau_C \left[ 0.86 - 0.79 \left(\log R_{\rm X} + 4.83 \right) \right]}{0.407 \left(B - V - 0.495 \right)^{0.325}} \right)^{1.767},
\label{eq:xray_ages}
\end{equation}
where $\tau_C$ is the convective overturn timescale as approximated by Equations \eqref{eq:t_conv1} and \eqref{eq:t_conv2}.  MH08 report that this relation holds, with a scatter of 0.25 in Rossby number, for X-ray activity levels $-7 < \log R_{\rm X} < -4$.  At higher levels of X-ray activity, there appears to be little correlation between X-ray activity and stellar rotation, and hence, age \citep[e.g.~][]{Pizzolato+Maggio+Micela+etal_2003}.  As for chromospheric activity, this relation requires $B-V > 0.5$, and is poorly calibrated for $B-V \gtrsim 1$.  X-ray activity measurements thus provide only upper limits to the ages of many SEEDS Moving Group targets.  For these extremely active targets, we assign a uniform probability distribution in age up to the maximum age (dependent upon spectral type) accessible to these age indicators.  

\subsection{Gyrochronology}
\label{subsec:gyro}

As F-type and later stars age, their rotation periods grow \citep{Kraft_1967, Skumanich_1972}.  This is believed to be due to their convective zones, which generate stellar magnetic fields, extending to the surface and coupling to the stellar wind \citep{Mestel_1968, Pinsonneault|Kawaler+Sofia+etal_1989}.  Stars more massive than mid-F spectral type have radiative envelopes and weak stellar winds; they hardly spin down at all \citep{Barnes_2003}.  Later-type stars with accurate cluster ages generally show one of two rotation patterns.  At young ages, a large fraction of stars (especially low-mass stars) are extremely fast rotators, forming the so-called {\it C}-sequence \citep{Barnes_2003}.  These fast rotators are believed to have their outer convective envelopes only weakly coupled to their inner radiative regions, resulting in inefficient angular momentum loss.  Older clusters lack these rapid rotators, which are believed to have transitioned onto the {\it I}-sequence, in which the star approaches solid-body rotation \citep{Barnes_2003}.

Young stars spend a variable amount of time on the rapidly rotating {\it C}-sequence before transitioning to the {\it I}-sequence, the duration of rapid rotation decreasing with increasing stellar mass.  This timescale varies from $\sim$300 Myr for early M stars to 0 for F stars \citep{Barnes_2003}.  Some stars appear to be on the {\it I}-sequence even at substantially younger ages, indicating that these timescales include substantial scatter.  We treat them as the youngest ages accessible to gyrochronology, lower bounds on our age constraints using these secondary criteria.  For simplicity, we use a parameterization linear in $B-V$ color, from 300 Myr at $B-V = 1.5$ (early M) to 0 at $B-V=0.5$ (late F).

For older stars on the {\it I}-sequence, color-dependent gyrochronology relations have been derived by \cite{Barnes_2007} and re-calibrated by MH08.  The relation is identical to Equation \eqref{eq:rhk_ages}, except that the rotation period is measured directly rather than inferred from chromospheric activity.  The gyrochronological age estimate becomes
\begin{equation}
\frac{\tau}{\rm Myr} \approx \left( \frac{\tau_{\rm rot}}{0.407 \left(B - V - 0.495 \right)^{0.325}} \right)^{1.767}~.
\label{eq:gyro}
\end{equation}
The scatter about this relation is very large at young ages \citep[$\sim$1 dex, ][]{Mamajek+Hillenbrand_2008}; in addition, it only applies to stars on the rotational {\it I}-sequence.  \cite{Barnes_2007} only applies such a result to stars rotating more slowly than the 100 Myr ``gyrochrone.''  We adopt a similar criterion by setting a floor on the gyrochronological age of 0 to 300 Myr depending on color, as described above, together with an overall floor of 100 Myr.  A star with a younger age according to Equation \eqref{eq:gyro} will be assigned a uniform probability distribution of ages up to the floor appropriate to its color.  

MH08 have measured a scatter about Equation \eqref{eq:gyro} of 0.05 dex for stars on the {\it I}-sequence, and recommend adding an additional $\sim$15\% ($\sim$0.06 dex) to account for systematic uncertainties in the cluster ages used for calibration.  We therefore adopt 0.8 dex as the age uncertainty for slow rotators.  

\subsection{Lithium Abundance}
\label{subsec:lithium}

The strength of lithium absorption lines declines as a star ages and burns its initial supply of the fragile element.  Stars with convective zones approaching the surface carry lithium down into the hotter interior where it is subsequently destroyed.  Unfortunately, other mixing processes complicate this picture, and the details of convection depend strongly on stellar mass.  

Lithium can be a problematic age indicator \citep{Zuckerman+Song_2004}, as its abundance is extremely sensitive to the stellar accretion history \citep{Baraffe+Chabrier_2010}, but abundant lithium is a reliable indicator of stellar youth \citep{Bildsten+Brown+Matzner+etal_1997}.  Extensive observations of open clusters do enable crude lithium age estimates for some stars \citep{Sestito+Randich_2005}, though for much of the SEEDS MG sample, lithium provides only upper limits.  There is a considerable scatter between coeval stars and a strong color dependence, and therefore, lithium is considered more reliable for dating young clusters \citep{Soderblom_2010}.

In general, lithium abundance is significantly more problematic as an age indicator for single stars than the activity and rotation measurements described above \citep{Soderblom_2010}.  In the notes for each individual star, we comment briefly on the consistency of lithium abundances with these other indicators (Section \ref{sec:obs_details}).

\subsection{Isochrone Dating}
\label{subsec:isochrone}

Isochrones in color-magnitude space are among the most reliable methods for dating coeval clusters and associations of stars \citep{Song+Zuckerman+Bessell_2003}.  Unfortunately, they are much less reliable for individual stars.  Isochrone dating fails to produce a robust peak in the probability distribution in a large fraction of field stars, and typically has uncertainties of $\gtrsim$1 Gyr even for those stars on which it is successful \citep{Takeda+Ford+Sills+etal_2007}.  In order for isochrone placement to have any value as an age indicator, a main-sequence star must have completed at least $\sim$1/3 of its life \citep{Soderblom_2010}.  This excludes most of the SEEDS Moving Group sample.  
In addition, any isochrone-based age analysis should marginalize over uncertainties in convection, composition, rotation, and atmospheric modeling, among other numerical and theoretical considerations.  We therefore do not attempt a full isochrone age analysis in this work.  However, isochrone ages can still provide an important check on ages estimated from other methods, and in particular, on the likelihood of a star's membership in a young moving group.  We therefore use the PARSEC stellar models \citep{Bressan+Marigo+Girardi+etal_2012} as a consistency check on the median ages we obtain by our full analysis (Section \ref{sec:BayesAges}).

A model of stellar structure, combined with a model atmosphere, predicts absolute magnitudes $M_i$ in a variety of bandpasses $i$.  Given observed (apparent) magnitudes $m_i$ in each band, we can write down the logarithmic likelihood of a model, together with a parallax $\varpi$ (in milliarcseconds), as
\begin{align}
-2 \ln {\cal L}({\rm mod}, \varpi) &= \sum_{{\rm bands}~i} \frac{\left(M_{i} + 5 \log_{10} 100/\varpi - m_{{\rm obs}, i} \right)^2}{\sigma^2_i} \nonumber \\
& \qquad + \frac{\left( \varpi - \varpi_{\rm Hip} \right)^2}{\sigma^2_{\varpi}}~.
\label{eq:iso_like}
\end{align}
We multiply Equation \eqref{eq:iso_like} by a prior in parallax prior equivalent to a uniform prior in space, $\varpi^{-4} d\varpi$, and marginalize over $\varpi$.  We adopt a Gaussian prior on [Fe/H] centered on the Solar value, with a standard deviation of 0.15 (40\% in metallicity).  This is nearly the same prior as that used by \cite{Nielsen+Liu+Wahhaj+etal_2013}, taken from the distribution of young FG dwarfs observed by \cite{Casagrande+Schonrich+Asplund+etal_2011}.  While this metallicity distribution should be appropriate for young stars, it is likely to systematically overestimate the metallicity (and photospheric opacity) of older stars.  Since stars brighten during their main-sequence lives, an overestimated metallicity would require an older age to compensate, and could produce large uncertainties in age determinations of several Gyr.

In an effort to be as uniform across the sample as possible, we restrict ourselves to the magnitudes measured by {\it Tycho} \citep{Hog+Fabricius+Makarov+etal_2000} and by 2MASS \citep{Cutri+Skrutskie+vanDyk+etal_2003}.  We do not attempt to marginalize over stellar mass and evolutionary rate in the color-magnitude diagram, both of which would be necessary for a full isochrone-based age analysis.  Stellar rotation, which can have a significant effect on evolutionary tracks and produce colors and luminosities that vary with viewing angle \citep{Ekstrom+Georgy+Eggenberger+etal_2012}, becomes another major uncertainty for more massive stars. 

\cite{Nordstrom+Mayor+Andersen+etal_2004} found that, for age probability distributions normalized over stellar mass, metallicity, and evolutionary rate in the color-magnitude diagram, 1-$\sigma$ confidence intervals corresponded roughly to a 60\% of the marginalized peak likelihood.  In our analysis, we adopt a more conservative threshold of $\Delta \ln {\cal L} = 1$, a ratio of $\sim$0.37.  Table \ref{tab:stellarages_full} includes the likelihood ratios; we comment on the stars with large discrepancies when we discuss the individual targets in Section \ref{sec:obs_details}.  In two cases, the isochrone checks lead us to reduce our estimated probability of moving group membership.

\begin{deluxetable*}{lcccccccr}
\tablewidth{0pt}
\vspace{-0.2truein}
\tablecaption{The SEEDS Moving Group Target List: Age Indicators}
\tablehead{
    \colhead{Name} &
    \colhead{Moving} &
    \colhead{Group} &
    \colhead{$\log R'_{\rm HK}$\tablenotemark{b}} &
    \colhead{$\log R_{\rm X}$\tablenotemark{c}} &
    \multicolumn{2}{c}{Li EW (m\AA)} &
    \colhead{$P_{\rm rot}$} &
    \colhead{Activity/Rotation} \\
    \colhead{} &
    \colhead{Group} &
    \colhead{References\tablenotemark{a}} &
    \colhead{} &
    \colhead{} &
    \colhead{Lit} &
    \colhead{APO} & 
    \colhead{(days)} &
    \colhead{References\tablenotemark{a}}
}
    
\startdata
HIP 544 & Her~Lya & 1, 2 & $-4.38$ & $-4.22$ & 75 & 92 & 6.23 & 13, 15, 16 \\
HIP 1134 & Columba & 3 & $-4.42$ & $-4.18$ & 99 & 128 & \ldots & 3, 15, 17, 18 \\
FK Psc & $\ldots$\tablenotemark{*} & 4/5 & \ldots & $-3.35$ & \ldots & \ldots & 7.7 & 19 \\
HIP 3589 & AB~Dor & 6 & \ldots & $-3.87$ & 199 & \ldots & \ldots & 11 \\
HIP 4979 & IC~2391 & 7 & \ldots & $-5.46$ & \ldots & \ldots & \ldots & \ldots \\
HIP 6869 & IC~2391 & 7 & $-4.76$ & $-4.89$ & 5 & 18 & \ldots & 17, 18 \\
HS Psc & AB~Dor & 8 & \ldots & $-3.08$ & 90 & \ldots & 1.09 & 20, 21 \\
HIP 10679 & $\beta$~Pic & 6 & $-4.37$ & $-3.84$\tablenotemark{B} & 160 & 168 & \ldots & 11, 17 \\
BD+30 397B & $\beta$~Pic & 9 & \ldots & $-2.55$\tablenotemark{B} & 110 & \ldots & \ldots & 11\\
HIP 11437 & $\beta$~Pic & 6 & \ldots & $-2.98$\tablenotemark{B} & 220 & 252 & 13.7 & 11, 21 \\
HIP 12545 & $\beta$~Pic & 6 & \ldots & $-2.98$\tablenotemark{B} & 450 & 436 & 1.25 & 11, 19 \\
HIP 12638 & AB~Dor & 6 & $-4.92$ & $-3.90$ & 158 & \ldots & \ldots & 11, 15 \\
HIP 12925 & Tuc-Hor & 3 & \ldots & $-4.26$ & 145 & 144 & \ldots & 3 \\
HIP 17248 & Columba & 3 & \ldots & $-3.36$ & \ldots & \ldots & \ldots & 3 \\
HIP 23362 & Columba & 3 & \ldots & $<$$-6.28$ & \ldots & \ldots & \ldots & \ldots \\
HIP 25486 & $\beta$~Pic & 6, 9 & $-4.27$ & $-3.46$ & 191 & 154 & \ldots & 11, 15 \\
HD 36869 & Columba & 3, 5 & \ldots & $-3.47$ & 204 & 210 & 1.31 & 3, 22 \\
HIP 29067 & Castor & 7, 10 & -4.43 & $<$$-4.48$ & 38 & \ldots & \ldots & 10, 15, 23, 24, 25 \\
HIP 30030 & \ldots\tablenotemark{*} & 5/9/11 & $-4.18$ & $-3.61$ & 170 & 164 & 1.15 & 11, 15, 22 \\
HIP 32104 & Columba & 3 & \ldots & $-5.57$ & \ldots & \ldots & \ldots & \ldots \\
V429 Gem & AB~Dor & 6 & $-4.2$ & $-3.37$ & 105 & 122 & 2.80 & 11, 19, 26 \\
HIP 37288 & Her~Lya & 2 & $-4.67$ & $<$$-4.74$ & 43 & \ldots & \ldots & 10, 23 \\
HIP 39896 & Columba & 5/7 & $-4.05$ & $-3.13$ & \ldots & 25 & 3.37 & 13, 27 \\
HIP 40774 & IC~2391 & 7 & $-4.45$ & $<$$-4.55$ & \ldots & 17 & \ldots & 28 \\
HIP 44526 & Castor & 7 & $-4.36$ & $-4.02$ & \ldots & 58 & 8.64 & 29, 30 \\
HIP 45383 & Castor & 10 & $-4.41$ & $-3.97$ & 9 & \ldots & \ldots & 10, 17, 23, 24, 25 \\
HIP 46843 & Columba\tablenotemark{*} & 5/7 & $-4.21$ & $-3.84$ & 176 & 188 & 5.38 & 10, 13, 31, 32 \\
HIP 50156 & Columba\tablenotemark{*} & 5/12 & $-3.96$ & $-3.39$ & \ldots & \ldots & 7.98 & 13, 33 \\
GJ 388 & Castor & 13 & $-4.17$ & $-3.10$ & \ldots & \ldots & 2.23 & 15, 34, 35 \\
HIP 50660 & IC~2391 & 7 & $-4.60$ & $<$$-4.35$ & \ldots & \ldots & \ldots & 28 \\
HIP 51317 & AB~Dor & 3, 5 & $-5.01$ & $-5.18$ & \ldots & \ldots & \ldots & 15  \\
HIP 53020 & Her~Lya & 2 & $-5.29$ & $<$$-4.36$ & \ldots & \ldots & \ldots & 15 \\
HIP 53486 & Castor & 7 & $-4.48$ & $-4.50$ & \ldots & 19 & 11.43 & 15, 30 \\
HD 95174 & $\beta$~Pic & 14 & \ldots & $<$$-4.54$\tablenotemark{B} & \ldots & 3 & \ldots & \ldots \\
HIP 54155 & Loc.~Ass. & 7 & $-4.35$ & $-3.63$ & 104 & 114 & \ldots & 10, 24, 25, 36 \\
TWA  2 & TW~Hya & 6 & \ldots & $-3.26$\tablenotemark{B} & 535 & \ldots & 4.86 & 11, 19 \\
TYC 3825-716-1 & AB~Dor & 14 & \ldots & $-3.28$ & \ldots & 36 & \ldots & \ldots \\
HIP 59280 & IC~2391 & 7, 10 & $-4.65$ & $-5.13$ & 26 & 18 & \ldots & 10, 15, 17 \\
TYC 4943-192-1 & AB~Dor & 8 & \ldots & $-3.45$ & \ldots & \ldots & \ldots & 8 \\
HIP 60661 & Loc.~Ass. & 7 & $-4.82$ & $<$$-4.33$ & \ldots & \ldots & \ldots & 13 \\
HIP 63317 & Loc.~Ass. & 13 & $-4.19$ & $-3.56$ & 94 & 106 & \ldots & 2, 13 \\
FH CVn & AB~Dor & 14 & \ldots & $-3.15$ & \ldots & \ldots & 2.17 & 14, 27 \\
HIP 66252 & IC~2391 & 7, 10 & $-3.89$ & $-3.12$ & 65 & 47 & 3.9 & 25, 33 \\
HIP 67412 & IC~2391 & 7 & $-4.64$ & $-5.00$ & \ldots & 15 & \ldots & 37 \\
HIP 73996 & UMa\tablenotemark{*} & 7/10 & $-4.38$ & $-5.33$ & \ldots & 20 & \ldots & 15 \\
HIP 78557 & Loc.~Ass. & 13 & $-4.20$ & $-4.60$ & 103 & \ldots & \ldots & 13 \\
HIP 82688 & AB~Dor & 5, 6 & $-4.29$ & $-4.18$ & 133 & 137 & \ldots & 11, 15 \\
HIP 83494 & Tuc-Hor\tablenotemark{*} & 3/5 & \ldots & $<$$-6.01$ & \ldots & \ldots & \ldots & \ldots \\
HIP 87579 & Castor & 13 & $-4.44$ & $-4.70$ & \ldots & \ldots & \ldots & 13, 17, 24, 25, 38 \\
HIP 87768 & Loc.~Ass. & 7 & $-4.27$ & $-4.72$ & 7 & \ldots & \ldots & 13, 24, 39 \\
HIP 91043 & Loc.~Ass. & 7 & $-4.21$ & $-3.30$ & 208 & \ldots & 1.34 & 13, 18, 40 \\
HIP 93580 & AB~Dor & 3, 5 & \ldots & $-5.21$ & \ldots & \ldots & \ldots & \ldots \\
BD+05 4576 & AB~Dor & 8 & \ldots & $-3.94$ & \ldots & \ldots & \ldots & \ldots     \\
HIP 102409 & $\beta$~Pic & 9 & $-4.11$ & $-2.86$ & 80 & \ldots & 4.85 & 11, 15, 19 \\
HD 201919 & AB~Dor & 6 & \ldots & $-3.49$ & 20 & \ldots & 4.92 & 11, 19 \\
HIP 107350 & Her~Lya & 1, 2 & $-4.42$\tablenotemark{MW} & $-4.39$ & 115 & 102 & 4.74 & 13, 31, 41 \\
TYC 2211-1309-1 & $\beta$~Pic & 4 & \ldots & $-3.11$ & $<$40 & \ldots & 0.476 & 19, 20 \\
HIP 111449 & Her~Lya & 2 & $-4.53$ & $-5.03$ & \ldots & \ldots & \ldots & 36, 42 \\
HIP 114066 & AB~Dor & 6 & \ldots & $-3.03$ & 30 & \ldots & 4.50 & 43, 44 \\
HIP 115162 & AB~Dor & 6 & $-4.22$ & $-4.22$ & 160 & 161 & \ldots & 25, 43 \\
BD$-$13 6424 & $\beta$~Pic & 6 & \ldots & $-3.05$ & 185 & 184 & 5.68 & 11, 19 \\
HIP 116805 & Columba & 3 & \ldots & $<$$-6.59$ & \ldots & \ldots & \ldots & \ldots
\enddata

\tablenotetext{a}{References: 1 \citep{Fuhrmann_2004}; 2 \citep{Lopez-Santiago+Montes+Crespo-Chacon+etal_2006}; 3 \citep{Zuckerman+Rhee+Song+etal_2011}; 4 \citep{Lepine+Simon_2009}; 5 \citep{Malo+Doyon+Lafreniere+etal_2013}; 6 \citep{Torres+Quast+Melo+etal_2008}; 7 \citep{Montes+Lopez-Santiago+Galvez+etal_2001}; 8 \citep{Schlieder+Lepine+Simon_2010}; 9 \citep{Zuckerman+Song_2004}; 10 \citep{Maldonado+Martinez-Arnaiz+Eiroa+etal_2010}; 11 \citep{daSilva+Torres+deLaReza+etal_2009}; 12 \citep{Schlieder+Lepine+Simon_2012a}; 13 \citep{Lopez-Santiago+Montes+Galvez-Ortiz+etal_2010}; 14 \citep{Schlieder+Lepine+Simon_2012b}; 15 \citep{Isaacson+Fischer_2010}; 16 \citep{Gaidos+Henry+Henry_2000}; 17 \citep{Wright+Marcy+Butler+etal_2004}; 18 \citep{White+Gabor+Hillenbrand_2007}; 19 \citep{Messina+Desidera+Tutatto+etal_2010}; 20 \citep{McCarthy+White_2012}; 21 \citep{Norton+Wheatley+West+etal_2007}; 22 \citep{Messina+Rodono+Guinan+etal_2001}; 23 \citep{Duncan+Vaughan+Wilson+etal_1991}; 24 \citep{Gray+Corbally+Garrison+etal_2003}; 25 \citep{Martinez-Arnaiz+Maldonado+Montes+etal_2010}; 26 \citep{Hernan-Obispo+Galvez-Ortiz+Anglada-Escude+etal_2010}; 27 \citep{Hartman+Bakos+Noyes+etal_2011}; 28 \citep{Pace_2013}; 29 \citep{Arriagada_2011}; 30 \citep{Strassmeier+Washuett+Granzer+etal_2000}; 31 \citep{Baliunas+Sokoloff+Soon_1996}; 32 \citep{Messina+Guinan+Lanza+etal_1999}; 33 \citep{Torres+Busko+Quast_1983}; 34 \citep{Cincunegui+Diaz+Mauas_2007}; 35 \citep{Hunt-Walker+Hilton+Kowalski+etal_2012}; 36 \citep{Schroder+Reiners+Schmitt_2009}; 37 \citep{Jenkins+Murgas+Rojo+etal_2011}; 38 \citep{Soderblom_1985}; 39 \citep{Favata+Barbera+Micela+etal_1993}; 40 \citep{Henry+Fekel+Hall_1995}; 41 \citep{Frasca+Ferrero+Marilli+etal_2000}; 42 \citep{Gray+Corbally+Garrison+etal_2006}; 43 \citep{Zuckerman+Song+Bessel_2004}; 44 \citep{Koen+Eyer_2002}}

\tablenotetext{b}{Values marked with `MW' are from multi-decade Mt.~Wilson measurements.}
\tablenotetext{c}{Values or approximate upper limits from the {\it ROSAT} All Sky Survey \citep{Voges+Aschenbach+Boller+etal_1999, Voges+Aschenbach+Boller+etal_2000, Hunsch+Schmitt+Sterzik+etal_1999}.  See Section \ref{subsec:xray_activity} for details.}
\tablenotetext{*}{References disagree on membership.  See section on individual stars for details.}
\label{tab:stellarages_full}
\end{deluxetable*}

\section{Bayesian Stellar Ages}
\label{sec:BayesAges}

The SEEDS moving group sample comes from many different associations; some of these are well-defined, while others are considered far less reliable.  Likewise, the confidence with which each target is identified as a moving group member varies considerably.  Most of the targets also have other age indicators, described in the previous section, which should be combined with the age inferred from moving group identification to produce the most reliable age estimate.  

We adopt a Bayesian approach to stellar ages, using as our prior a flat age distribution out to 10 Gyr (appropriate to the local disk) or to the star's main sequence lifetime, and derive posterior probability distributions using age indicators and moving group memberships.  A slightly different star formation history, like the enhancement by a factor of 1.5 from 1 and 4 Gyr before the present \citep{Girardi+Groenewegen+Hatziminaoglou+etal_2005}, would have little effect on our results.  The resulting posterior probability distributions on age are suitable inputs to a statistical analysis of exoplanet frequencies and properties.  

The likelihood function ${\cal L}$ is difficult to write down.  If moving group membership and the stellar age indicators were all independent of one another, then ${\cal L}$ would simply be the product of the probability of group membership and the probability distribution inferred for each indicator.  However, moving group membership is often assigned, at least partially, on the basis of stellar activity.  Furthermore, indicators of stellar youth physically arise from the interplay of rotation and convection: chromospheric activity, rotation period, and coronal activity are not independent.  

Many authors have performed detailed analyses of moving groups, assigning membership probabilities to each proposed member.  We generally defer to these probabilities and adopt the moving group age distribution ${\cal L}_{\rm MG} (\tau)$ weighted by the membership probability $P_{\rm MG}$.  We approximate the moving group age likelihood function as a Gaussian with the confidence intervals described in Section \ref{sec:moving_groups} representing its median age $\pm 2 \sigma$.  The other age indicators, described in Section \ref{sec:otherindicators} and listed in Table \ref{tab:stellarages_full}, complement the group age for stars with uncertain membership or which belong to less well-defined associations.  We denote the likelihood function based solely on these single star indicators by ${\cal L}(\tau | {\rm indic}$; the total likelihood function is simply
\begin{equation}
{\cal L} ( \tau ) = P_{\rm MG} {\cal L}_{\rm MG} (\tau ) + \left( 1 - P_{\rm MG} \right) {\cal L} (\tau | {\rm indic} )~. \label{eq:prob_age}
\end{equation}
Equation \eqref{eq:prob_age} assumes the age derived from secondary indicators and proposed moving group membership to be independent, which could be problematic.  In this analysis, it is not a major problem, as most of our stars are either reliably associated with a moving group or have no kinematic age that we trust.  

As described in the previous section, MH08 find the best results for the activity age by first using chromospheric and coronal activity to estimate the Rossby number, and then using gyrochronology to estimate stellar age.  We therefore treat coronal and chromospheric activity as independent measurements of Rossby number Ro.  In practice, the scatter in Ro as estimated by X-ray activity is $\sim$1.5 times as large as that estimated by $R'_{\rm HK}$ (MH08) over the applicable activity regimes, so this approximation has little practical effect.  The situation is dramatically better for the (one) object with multi-decade Mt.~Wilson chromospheric data.  

It is more difficult to estimate the covariance between stellar age as estimated from activity via the Rossby number and that inferred directly from a rotation period.  The latter estimator, being more direct, has a smaller scatter reported in MH08 ($\sim$0.05 dex) than the activity-rotation age ($\sim$0.1 dex from binaries, $\sim$0.2 dex from clusters) for stars on the {\it I} rotational sequence.  As the SEEDS sample painfully illustrates, however, this does not include all variation in rotation at a common age.  Our slowest rotator, HIP 11437, has a gyrochonological age of $\sim$500 Myr, but is reliably identified with the $\beta$ Pic moving group.  The star might still be contracting onto the main sequence, or it could simply be an outlier.  MH08 also omitted two anomalously slow rotators in the Pleiades from their analysis.

With the above caveat, we note that assuming the age indicators to be independent makes little difference; the scatter in the period-age relation is much smaller than in the activity-age relations.  We therefore simply set 0.05 dex as the floor in the uncertainty and add 0.06 dex to the error estimated from activity and rotation to account for systematic uncertainties in the cluster ages used to calibrate the relations (MH08).  MH08 only used the slow, {\it I}-sequence rotators to derive their gyrochronological ages; we therefore add the range of time spent on the rapidly rotating {\it C}-sequence, $\sim$100--300 Myr, to the age distributions (see Section~\ref{subsec:gyro}).

All of the activity/period/age relations have a strong color dependency, with later spectral types spinning down more rapidly after reaching the {\it I}-sequence.  Spectral types earlier than late F, with $B-V$ colors $\lesssim$0.5, never reach the {\it I}-sequence.  They never achieve the deep convective zone and strong dynamo necessary to drive a magnetized wind, and rotate rapidly throughout their main sequence lifetimes.  For such stars in the SEEDS Moving Group sample, we have little choice but to use a flat probability distribution out to the star's main sequence lifetime.  We also note that the relations derived in MH08 were only tested for FGK stars, with $0.5 \lesssim B-V \lesssim 0.9$.  Many of our targets are late K and M stars with colors as red as $B-V \sim 1.5$.  The basic rotation/activity/age relation should continue to hold for these stars, albeit with larger uncertainties.  We therefore continue to apply the relationships, noting that the $\sim$300 Myr timescale to reach the {\it I}-sequence adds a large spread to the derived ages.  

Many of our targets are relatively faint and, as such, have poor {\it Tycho} measurements of $B-V$.  We therefore combine the {\it Tycho} colors with a $B-V$ color estimated from $V-K$, with $V$ measured from {\it Tycho} (transformed to Johnson) and $K$ measured from 2MASS, using Table 5 of \cite{Pecaut+Mamajek_2013}.  We find that, in order to reproduce the scatter of {\it Tycho} colors using converted $V-K$ magnitudes, we need to add an empirical error of $\sim$0.03 mag to the interpolated result.  We then combine the two estimates of $B-V$.  This gives a median final uncertainty $\sigma_{B-V} = 0.018$ mag, and $\sigma_{B-V} < 0.05$ mag for all but one star.

The very old and very young tails of the probability distribution are important (and extremely difficult) to model properly.  Several stars in our sample make this all too evident, with disturbingly discrepant kinematic and activity ages.  This will become much more of a problem as high-contrast surveys begin to report larger numbers of detections, and the properties of individual exoplanet host stars are subjected to higher scrutiny.  For now, we note that authors estimating ages from clusters routinely throw out a few percent of their stars as pathological cases \citep[e.g.~][]{Mamajek+Hillenbrand_2008}.  We therefore account for these long tails, at least qualitatively, by giving each target not definitively associated with a moving group a 5\% probability of being pathological, with utterly uninformative age indicators.  More work on large samples of young stars should help to constrain the intrinsic scatter in activity and rotation at a common age.

Table \ref{tab:ages} summarizes the posterior probability distributions on age for all of the SEEDS MG targets.  The third column lists the adopted membership probability in the indicated moving group (see Section \ref{sec:obs_details} for details on individual stars), with `\ldots' for those groups that we do not consider to be sufficiently well-defined to provide secure age estimates.  The fourth and fifth columns list the 5\% and 95\% edges of the age probability distribution exclusively on the secondary age indicators, while the final three columns list the final 5\%, 50\%, and 95\% ages based on all available information.  For those stars without any age constraints beyond their finite main sequence lifetimes, we list `\ldots' in columns 4-8.

\begin{figure}
\includegraphics[width=\linewidth]{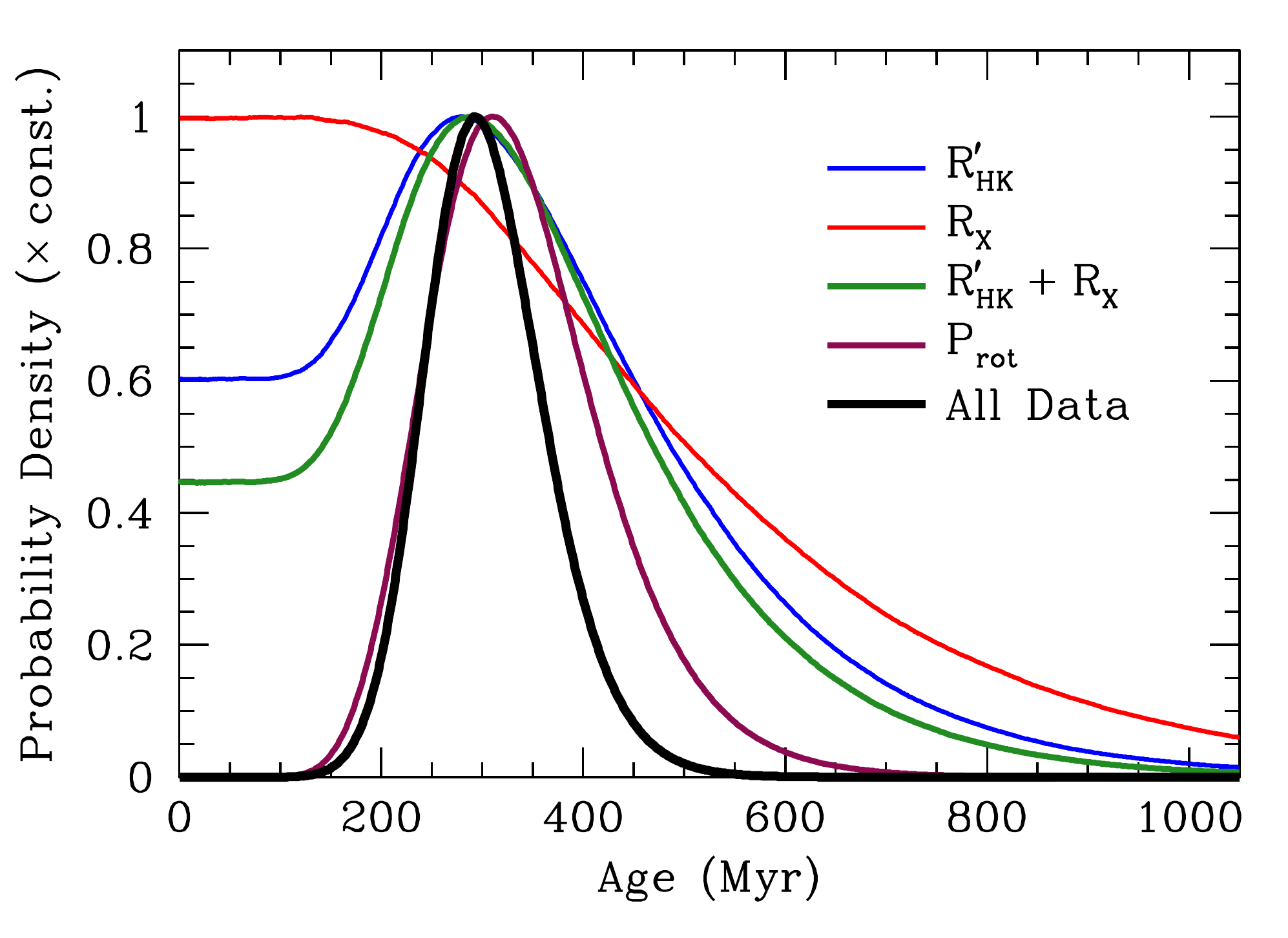}
\caption{The calculation of an age probability distribution for a target, HIP 107350, without a reliable moving group age.  HIP 107350 has an exceptional array of secondary age indicators, which enable a good constraint on its age.  Most other stars without kinematic ages have much broader posterior probability distributions.}
\label{fig:age_demo}
\end{figure}

Figure \ref{fig:age_demo} demonstrates our age determination method for HIP 107350.  This star lacks a secure moving group age, but has an exceptional array of secondary age indicators, including a measured rotation period and multi-decade Mt.~Wilson chromospheric activity measurements.  As a G0 star, HIP 107350 represents the best possible case for the use of secondary age indicators.  

\begin{deluxetable*}{lcccccccr}
\tablewidth{0pt}
\tablecaption{The SEEDS Targets' Ages}
\tablehead{
    \colhead{Name} &
    \colhead{Moving} &
    \colhead{Membership} &
    \multicolumn{2}{c}{No Group Data (Myr)\tablenotemark{b}} &
    \multicolumn{3}{c}{All Data (Myr)} &
    \colhead{$\Delta \ln {\cal L}$} \\
    \colhead{} &
    \colhead{Group} &
    \colhead{Prob.~(\%)\tablenotemark{a}} &
    \colhead{5\%} &
    \colhead{95\%} &
    \colhead{5\%} &
    \colhead{50\%} &
    \colhead{95\%} &
    \colhead{}
}

\startdata
HIP 544 & Her Lya & \ldots & 190 & 370 & 190 & 270 & 370 & 0.1 \\ 
 HIP 1134 & Columba & 95+ & 24 & 940 & 20 & 30 & 54 & 0.38 \\ 
 FK Psc & $\beta$ Pic & 20 & 190 & 460 & 18 & 290 & 450 & \ldots  \\ 
 HIP 3589 & AB Dor & 95+ & 11 & 480 & 110 & 130 & 150 & 8.3 \\ 
 HIP 4979 & IC 2391 & \ldots & \ldots & \ldots & \ldots & \ldots & \ldots & 6.4 \\ 
 HIP 6869 & IC 2391 & \ldots & 790 & 3300 & 790 & 1400 & 3300 & \ldots  \\ 
 HS Psc & AB Dor & 95+ & 5 & 180 & 110 & 130 & 150 & \ldots  \\ 
 HIP 10679 & $\beta$ Pic & 95+ & 620 & 8300 & 16 & 21 & 28 & 0.085 \\ 
 BD+30 397B & $\beta$ Pic & 95+ & 820 & 9200 & 16 & 21 & 28 & \ldots  \\ 
 HIP 11437 & $\beta$ Pic & 95+ & 520 & 1100 & 16 & 21 & 29 & \ldots  \\ 
 HIP 12545 & $\beta$ Pic & 95+ & 5 & 3900 & 16 & 21 & 24 & \ldots  \\ 
 HIP 12638 & AB Dor & 95+ & 1800 & 4500 & 110 & 130 & 170 & 0.27 \\ 
 HIP 12925 & Tuc-Hor & 95+ & 16 & 750 & 16 & 180 & 750 & 0.3 \\ 
 HIP 17248 & Columba & 95+ & 27 & 1100 & 20 & 30 & 54 & \ldots  \\ 
 HIP 23362 & Columba & 95+ & \ldots & \ldots & 20 & 30 & 54 & 1 \\ 
 HIP 25486 & $\beta$ Pic & 95+ & 8 & 250 & 16 & 21 & 24 & 2.2 \\ 
 HD 36869 & Columba & 95+ & 26 & 120 & 20 & 30 & 50 & 0.17 \\ 
 HIP 29067 & Castor & \ldots & 1100 & 9200 & 1100 & 5100 & 9200 & \ldots  \\ 
 HIP 30030 & Columba & 95+ & 27 & 120 & 20 & 30 & 50 & 0.072 \\ 
 HIP 32104 & Columba & 95+ & \ldots & \ldots & 20 & 30 & 57 & 0.023 \\ 
 V429 Gem & AB Dor & 95+ & 45 & 240 & 110 & 130 & 150 & \ldots  \\ 
 HIP 37288 & Her Lya & \ldots & 1500 & 9300 & 1500 & 5300 & 9300 & \ldots  \\ 
 HIP 39896 & Columba & 50 & 57 & 290 & 22 & 49 & 280 & \ldots  \\ 
 HIP 40774 & IC 2391 & \ldots & 1200 & 9300 & 1200 & 5100 & 9300 & \ldots  \\ 
 HIP 44526 & Castor & \ldots & 310 & 570 & 310 & 430 & 570 & \ldots  \\ 
 HIP 45383 & Castor & \ldots & 460 & 2000 & 460 & 970 & 2000 & \ldots  \\ 
 HIP 46843 & Columba & \ldots & 160 & 310 & 160 & 240 & 310 & 0.61 \\ 
 HIP 50156 & Columba & 80 & 200 & 480 & 21 & 33 & 410 & \ldots  \\ 
 GJ 388 & Castor & \ldots & 40 & 330 & 40 & 190 & 330 & \ldots  \\ 
 HIP 50660 & IC 2391 & \ldots & 980 & 9200 & 980 & 4900 & 9200 & 0.83 \\ 
 HIP 51317 & AB Dor & 95+ & 1800 & 9300 & 110 & 130 & 170 & \ldots  \\ 
 HIP 53020 & Her Lya & \ldots & 1100 & 9200 & 1100 & 5000 & 9200 & \ldots  \\ 
 HIP 53486 & Castor & \ldots & 540 & 990 & 540 & 720 & 990 & \ldots  \\ 
 HD 95174 & $\beta$ Pic & 10 & 1200 & 9200 & 20 & 4600 & 9200 & \ldots  \\ 
 HIP 54155 & Loc.~Ass & \ldots & 28 & 990 & 28 & 300 & 990 & 0.4 \\ 
 TWA 2 & TW Hya & 95+ & 810 & 9200 & 5 & 10 & 19 & \ldots  \\ 
 TYC 3825-716-1 & AB Dor & \ldots & 27 & 1100 & 27 & 280 & 1100 & \ldots  \\ 
 HIP 59280 & IC 2391 & \ldots & 670 & 2500 & 670 & 1300 & 2500 & 0.48 \\ 
 TYC 4943-192-1 & AB Dor & 80 & 27 & 1100 & 110 & 130 & 450 & \ldots  \\ 
 HIP 60661 & Loc.~Ass & \ldots & 980 & 9200 & 980 & 5000 & 9200 & \ldots  \\ 
 HIP 63317 & Loc.~Ass & \ldots & 25 & 870 & 25 & 260 & 870 & 0.0075 \\ 
 FH CVn & AB Dor & 40 & 32 & 230 & 40 & 130 & 220 & \ldots  \\ 
 HIP 66252 & IC 2391 & \ldots & 73 & 270 & 73 & 170 & 270 & \ldots  \\ 
 HIP 67412 & IC 2391 & \ldots & 970 & 3100 & 970 & 1700 & 3100 & 0.46 \\ 
 HIP 73996 & UMa & \ldots & \ldots & \ldots & 280 & 2700 & 5100 & 2.6 \\ 
 HIP 78557 & Loc.~Ass & \ldots & 57 & 1400 & 57 & 400 & 1400 & 0.2 \\ 
 HIP 82688 & AB Dor & 95+ & 16 & 530 & 110 & 130 & 150 & 0.41 \\ 
 HIP 83494 & Tuc-Hor & \ldots & \ldots & \ldots & \ldots & \ldots & \ldots & 4.3 \\ 
 HIP 87579 & Castor & \ldots & 240 & 3200 & 240 & 1200 & 3200 & \ldots  \\ 
 HIP 87768 & Loc.~Ass & \ldots & 270 & 3100 & 270 & 1200 & 3100 & \ldots  \\ 
 HIP 91043 & Loc.~Ass & \ldots & 12 & 330 & 12 & 130 & 330 & 11 \\ 
 HIP 93580 & AB Dor & 80 & \ldots & \ldots & 110 & 130 & 1800 & 2.8 \\ 
 BD+05 4576 & AB Dor & 40 & 27 & 1100 & 46 & 140 & 800 & \ldots  \\ 
 HIP 102409 & $\beta$ Pic & 95+ & 92 & 360 & 16 & 21 & 28 & \ldots  \\ 
 HD 201919 & AB Dor & 95+ & 110 & 290 & 110 & 130 & 150 & \ldots  \\ 
 HIP 107350 & Her Lya & \ldots & 250 & 440 & 250 & 340 & 440 & 0.49 \\ 
 TYC 2211-1309-1 & $\beta$ Pic & 50 & 6 & 220 & 13 & 22 & 190 & \ldots  \\ 
 HIP 111449 & Her Lya & \ldots & \ldots & \ldots & \ldots & \ldots & \ldots & 0.9 \\ 
 HIP 114066 & AB Dor & 95+ & 84 & 340 & 110 & 130 & 150 & \ldots  \\ 
 HIP 115162 & AB Dor & 95+ & 26 & 1400 & 110 & 130 & 150 & 0.078 \\ 
 BD$-$13 6424 & $\beta$ Pic & 95+ & 120 & 390 & 16 & 21 & 28 & \ldots  \\ 
 HIP 116805 & Columba & 30 & \ldots & \ldots & 21 & 130 & 440 & 1.6 
\enddata

\tablenotetext{a}{High-confidence classifications from, e.g., \cite{Torres+Quast+Melo+etal_2008} and \cite{Malo+Doyon+Lafreniere+etal_2013}, including the corresponding web tool BANYAN.  See notes on individual objects for more doubtful classifications.}
\tablenotetext{b}{An entry of `\ldots' indicates that the star is too blue for the activity/rotation/age relations to apply, and that its age probability distribution is therefore uniform out to 10 Gyr or its main sequence lifespan.}
\label{tab:ages}
\end{deluxetable*}

\section{Observations and Data Reduction} 
\label{sec:data_reduction}

Table \ref{tab:observing_log} lists all of the SEEDS Moving Groups targets and observations through May of 2013.  All observations were made using the HiCIAO instrument \citep{Suzuki+Kudo+Hashimoto+etal_2010} and AO188 \citep{Hayano+Takami+Guyon+etal_2008} on the Subaru telescope, and nearly all were made in the $H$-band.  As with many other high-contrast imaging surveys \citep[e.g.][]{Lafreniere+Doyon+Marois+etal_2007, Vigan+Patience+Marois+etal_2012}, the $H$-band was chosen due to both the AO performance and the relative brightness of the expected companions.  A typical observation sequence consisted of target acquisition, AO tuning, and acquisition of photometric reference frames, followed by the main, saturated science data taken in pupil-tracking ADI mode.  Including all overheads, $\sim$1-1.5 hours of telescope time were spent on a typical object.  

All of our data were taken in ADI mode and reduced using the ACORNS-ADI software package. The software and data reduction process are described in detail in \cite{Brandt+McElwain+Turner+etal_2013}; we therefore give only a brief summary here.  The source code is freely available for download at \verb|http://www.github.com/t-brandt/acorns-adi|.

For each sequence of images, we calibrate the data, register the frames, subtract the stellar PSF using the LOCI algorithm \citep{Lafreniere+Marois+Doyon+etal_2007}, and combine the image sequence using an adaptive trimmed mean.  Calibration consists of the usual flat-fielding and bad pixel masking, together with an algorithm to suppress correlated read noise in HiCIAO's H2RG detector.  We then correct for field distortion by comparing observations of globular clusters made with HiCIAO and with the {\it Hubble Space Telescope}.  We register the frames in each ADI sequence using templates of saturated PSFs built from thousands of images of dozens of stars.  This registration technique is accurate to $\sim$0.3 HiCIAO pixels, or 3 mas, under good observing conditions.  We then set the absolute centroid of an image sequence by visual inspection.

ACORNS-ADI includes several algorithms to model and subtract the stellar PSF.  In this work, we exclusively use the LOCI algorithm \citep{Lafreniere+Marois+Doyon+etal_2007} due to its speed, simplicity, and Gaussian residuals.  As our fiducial LOCI parameters, we use an angular protection zone of 0.7 times the PSF full width at half maximum (FWHM), and optimization zones 200 PSF footprints in area.  Our subtraction regions vary in size from a few PSF footprints at small separations to a few tens of footprints several arcseconds from the central star. HiCIAO data is oversampled in the $H$-band, with a typical FWHM of 6 pixels.  We limit the number of LOCI comparison frames to avoid solving an under-constrained system and suppressing more companion flux than necessary---in the limit of an equal number of pixels and comparison frames, flux (and noise) suppression would be perfect.  The final contrast of an ADI reduction with LOCI is a concave function of the number of comparison frames used for PSF modeling and subtraction, with a broad peak at $\sim$80 frames.  We therefore treat large data sets as a number of smaller data sets (with every $n$-th frame), reduce each of these small data sets separately using ACORNS-ADI, and then average the results to produce a map of residual intensity.  

We calibrate the partial subtraction in LOCI using the procedure described in \cite{Brandt+McElwain+Turner+etal_2013}.  We also include the much smaller effects of field rotation during each individual exposure and uncertainties in image registration, and approximate degradation in the AO performance with separation from the guide star by
\begin{equation}
{\rm SR} \propto \exp \left[ -\left( \frac{\Delta \theta}{\theta_0} \right)^{5/3} \right]~,
\end{equation}
where SR is the Strehl ratio, proportional to a point source's peak intensity, and we use an isoplanatic angle $\theta_0 = 30''$ \citep{Minowa+Hayano+Watanabe+etal_2010}.  These are all small corrections for our data, generally a few percent within $\sim$5$''$ of the central star.  Finally, we convolve the map of residual intensity with a circular aperture, normalize by the azimuthal standard deviation in residual intensity, and search for $5.5\sigma$ outliers.  We perform photometric calibrations using unsaturated reference frames taken before, after, and sometimes during an ADI sequence, and normalize to the central star's $H$-band magnitude in the 2MASS catalog \citep{Cutri+Skrutskie+vanDyk+etal_2003}.  ACORNS-ADI produces 2D contrast maps.  We azimuthally average these maps to obtain the contrasts reported in Table \ref{tab:contrast_curves}. 

We follow up companion candidates ($5.5\sigma$ detections), typically after $\sim$1 year, to test for physical association.  The SEEDS Moving Groups targets are almost all within 50 pc, with proper motions of up to 1$''$/yr.  A physically unrelated background object will thus move by an easily detectable amount, while a bound companion will remain in nearly the same position relative to its parent star.  None of our faint, substellar companion candidates thus far have passed the ``common proper motion test,'' though we have detected several low-mass stellar companions, and a few substellar candidates remain to be followed up.  Table \ref{tab:binaries} summarizes the new stellar companions to the MG targets, one of which does not yet have a second epoch to confirm common proper motion (though very close chance alignments of such bright stars at the targets' Galactic coordinates are unlikely).  Unsurprisingly, the frequency of background objects increases sharply towards the Galactic plane.

\begin{deluxetable}{lcccr}
\tablewidth{0pt}
\tablecaption{Newly Discovered Stellar Companions}
\tablehead{
    \colhead{Star} &
    \colhead{MJD} &
    \colhead{Separation} &
    \colhead{P.A.} &
    \colhead{$H$-band} \\
    \colhead{} &
    \colhead{$+$55,000} &
    \colhead{(arcsec)} &
    \colhead{(deg)} &
    \colhead{Contrast} 
}
    
\startdata
HIP 6869\tablenotemark{a}   &  137 & $0.444 \pm 0.005$ & $269.1 \pm 0.6$ & 100 \\
HIP 12925  &  927 & $1.893 \pm 0.005$ & $252.9 \pm 0.2$ &  13 \\
HIP 39896  &  920 & $0.252 \pm 0.005$ &  $81 \pm 1$ & 6.4 \\
HIP 45383  &  646 & $0.741 \pm 0.005$ &  $45.9 \pm 0.4$ &   4.1 \\
HIP 60661  &  219 & $1.92 \pm 0.01$ & $107.5 \pm 0.3$ &   5.7 \\
HIP 78557  & 1116 & $0.565 \pm 0.005$ & $180.7 \pm 0.5$ & 190 \\
HIP 82688  &  705 & $3.811 \pm 0.005$ &  $58.3 \pm 0.1$ &  41
\enddata

\tablenotetext{a}{Common proper motion to be confirmed.}

\label{tab:binaries} 
\end{deluxetable}

\section{Notes on Individual Stars}
\label{sec:obs_details}

In this section, where appropriate, we provide details on each individual target.  This includes both stellar properties (particularly age indicators) and noteworthy aspects of the SEEDS observations.  We order the objects by right ascension.

{\it HIP 544 (= HD 166 = GJ 5)}---This K0 star is a proposed member of the Hercules-Lyra association.  It does have extensive secondary indicators, enabling a reasonable age estimate.  SEEDS images do not detect any companion candidates within $8.\!\!''5$ ($\sim$110 AU projected).

{\it HIP 1134 (= HD 984)}---This late F star is considered to be a reliable member of the Columba moving group, and its abundant lithium and strong activity are consistent with a young age.  SEEDS images detect no companion candidates within $7.\!\!''5$ ($\sim$350 AU projected). 

{\it FK Psc (= TYC 1186-706-1)}---The moving group membership of this K7 star is disputed.  \cite{Lepine+Simon_2009} propose membership in $\beta$ Pic, while \cite{Malo+Doyon+Lafreniere+etal_2013} find it to be a field star with $\sim$55\% confidence.  SEEDS observations have resolved it as a binary with a separation of $1.\!\!''7$ and a flux ratio of $\sim$2 in $H$.  These results make the moving group analyses much more difficult to interpret, and we do not consider FK Psc to be reliably associated with any of the groups discussed in this paper.  We follow \cite{Malo+Doyon+Lafreniere+etal_2013} in placing a $\sim$20\% probability on $\beta$ Pic membership.  We also consider the kinematic distance given in Table \ref{tab:stellarprops} to be unreliable, making it difficult to interpret sensitivity limits.  SEEDS observed FK Psc under poor conditions; due to this and its extremely uncertain age, FK Psc should probably be excluded from statistical analyses.  

{\it HIP 3589 (= HD 4277)}---This late F star has been classified as a member of AB Dor with high confidence by, e.g., \cite{Torres+Quast+Melo+etal_2008}.  The star does show a strong discrepancy between the moving group age and the isochrone likelihood, with the isochrone analysis showing strong peaks at $\sim$20 Myr and $\sim$5 Gyr.  The old age, however, is extremely inconsistent with HIP 3589's youth indicators.  HIP 3589 has a neighbor at a separation of $3.\!\!''0$ with an $H$-band flux $\sim$10\% that of the primary; however, SEEDS observations indicate that this star is not bound to HIP 3589.  SEEDS images did not detect any other companion candidates within $7.\!\!''5$, $\sim$400 AU projected.  

{\it HIP 4979 (= HD 6288A)}---This early F star has been proposed to be a member of the IC 2391 supercluster.  Unfortunately, its early spectral type renders secondary age indicators of little value, and the star is extremely difficult to date reliably.  Our adopted age probability distribution is uniform out to HIP 4979's main sequence lifetime of $\sim$5 Gyr.  SEEDS images detect no companion candidates, apart from a marginal, 5.7$\sigma$ source at a separation $({\rm E},\,{\rm N}) = (-3.\!\!''94,\,6.\!\!''65)$, a projected distance of just under 500 AU.  Follow-up observations with somewhat less integration time detected nothing at this position, but did detect an even more marginal source ($\sim$$4\sigma$) near the candidate's expected background position.  

{\it HIP 6869 (= HD 8941)}---This F8 star has been proposed to be a member of the IC 2391 supercluster.  Like HIP 4979, it is too blue to apply the age relations described in this paper, and we adopt a uniform probability distribution in age.  HIP 6869 is a close binary, with an angular separation of $0.\!\!''44$ and an $H$-band contrast of $\sim$100. At HIP 6869's distance, its companion has an absolute $H$-band magnitude of $\sim$7, consistent with a mid-M spectral type. No other companion candidates were detected in high-contrast imaging.  We have not yet followed up the star to confirm its companion's common proper motion, though a close chance alignment of such a bright star at $(l,\,b) = (135^\circ,\,-45^\circ)$ is unlikely.

{\it HS Psc}---This mid-K star was first proposed as a candidate member of AB Dor by \cite{Schlieder+Lepine+Simon_2010}.  \cite{Malo+Doyon+Lafreniere+etal_2013} confirmed this categorization, placing it in AB Dor with 98\% confidence.  SEEDS imaging detected a 5.9 $\sigma$ point source at a separation of $({\rm E},\,{\rm N}) = (2.\!\!''85,\,4.\!\!''04)$, a projected distance of just under 200 AU assuming the kinematic distance \cite{Schlieder+Lepine+Simon_2010} derived assuming membership in AB Dor.  Follow-up images failed to detect any point source, making it a likely statistical fluctuation.

{\it HIP 10679 (= BD 14082B)}---This early G star is in a binary system with the early F star HIP 10680, separated from its companion by 14$''$.  It is considered to be a well-established member of the $\beta$ Pic moving group \citep[e.g.][]{Torres+Quast+Melo+etal_2008, Malo+Doyon+Lafreniere+etal_2013}.  SEEDS did not detect any companion candidates within a projected separation of $7.\!\!''5$ ($\approx$210 AU).  

{\it BD$+$30 397B}---This M0 star is the binary companion of the K7 star HIP 11437; both are reliably identified with the $\beta$ Pic moving group.  SEEDS images detected no companion candidates within $8.\!\!''5$ ($\sim$425 AU projected).  

{\it HIP 11437}---Together with its companion BD$+$30 397B, this K6 star is reliably identified with the $\beta$ Pic moving group.  It also represents a pathological case of exceptionally slow rotation (as measured from SuperWASP periodicity): HIP11437's gyrochronological age is $\sim$500 Myr.  Its 14 day period is the longest in the SEEDS moving group sample.  SEEDS did not detect any companion candidates within $7.\!\!''5$ ($\sim$300 AU projected).  

{\it HIP 12545}---This K6 star is considered a well-established member of the $\beta$ Pic moving group.  \cite{Malo+Doyon+Lafreniere+etal_2013} found its photometry and radial velocity to be slightly more consistent with Columba, but due in large part to its exceptionally fast rotation and vigorous activity, they did not dispute the traditional association with $\beta$ Pic \citep{Zuckerman+Song_2004, Torres+Quast+Melo+etal_2008}.  SEEDS images do not detect any companion candidates within $8.\!\!''5$ ($\sim$350 AU projected).  

{\it HIP 12638 (= HD 16760)}---This G2 star is a well-established member of the AB Dor moving group, and is known to host a companion, which was reported as a substellar object ($M \sin i \sim 14 M_J$) on a 1.3 year orbit \citep{Bouchy+Hebrard+Udry+etal_2009, Sato+Fischer+Ida+etal_2009}.  This companion was directly imaged on an almost face-on orbit, probably indicating that the companion has a stellar mass \citep{Evans+Ireland+Kraus+etal_2012}.  Despite its very modest contrast, the companion has an angular separation of just $\sim$$0.\!\!''026$, less than the width of the $H$-band Subaru PSF; it was imaged using aperture-masking interferometry.  SEEDS images detect no companion candidates within $7''$ ($\sim$320 AU projected).

{\it HIP 12925 (= HD 17250)}---This F8 star has been reliably classified in the Tuc-Hor association \citep{Zuckerman+Rhee+Song+etal_2011, Malo+Doyon+Lafreniere+etal_2013}.  SEEDS imaging detects a companion candidate with an $H$-band flux ratio of $\sim$13 and a separation of $1.\!\!''9$ ($\sim$100 AU projected); there are no other companion candidates within $7.\!\!''5$ ($\sim$400 AU projected).  Archival images from Keck/NIRC2 confirm this candidate to be HIP 12925's stellar companion.  

{\it HIP 17248}---This M0 star is considered to be a reliable member of the Columba moving group \citep{Zuckerman+Rhee+Song+etal_2011, Malo+Doyon+Lafreniere+etal_2013}.  SEEDS images detected five candidate companions within $7''$, with $H$-band contrasts ranging from $\sim$$10^4$ to $\sim$$2 \times 10^5$, and separations ranging from $\sim$$3''$ to $\sim$$6.\!\!''5$.  The star is less than a degree from the Galactic plane, making the density of background objects high.  Indeed, all but one of our companion candidates are clearly visible as background objects in HST/NICMOS imaging from 2005.  The final candidate, at a separation $({\rm E},\,{\rm N}) = (-2.\!\!''85,\,0.\!\!''72)$ in our images from November 2012, also appears to be in its expected background position in the archival HST/NICMOS data, albeit at a modest signal-to-noise ratio.  

{\it HIP 23362 (= HD 32309)}---This late B star is a secure member of the Columba moving group.  At an age of 30 Myr for the group, the isochrone fit is modestly discrepant; however, the isochrone analysis produces a very broad peak in the likelihood centered at $\sim$60 Myr.  Primarily part of the SEEDS high-mass sample, we include it here for completeness.  SEEDS images detect two companion candidates which are currently awaiting follow-up observations.

{\it HIP 25486 (= HD 35850)}---This F8 star is a well-established member of $\beta$ Pic; its high activity and abundant lithium confirm its youth.  The large discrepancy with the isochrone likelihood at 20 Myr is simply because the likelihood increases very sharply towards $\sim$25--30 Myr, and does not call the $\beta$ Pic identification into question.  SEEDS images do not detect any companion candidates within $7.\!\!''5$ ($\sim$200 AU projected).  

{\it HD 36869}---This G3 star is a likely member of the Columba moving group, but lacks a {\it Hipparcos} parallax.  Though it is absent from the large Bayesian analysis of \cite{Malo+Doyon+Lafreniere+etal_2013}, BANYAN gives a membership probability of more than 95\%.  HD 36869's {\it Tycho} parallax \citep{Hog+Fabricius+Makarov+etal_2000} is far below the distance inferred from its magnitude and spectral type \citep[59 pc,~][]{Zuckerman+Rhee+Song+etal_2011}; we consider the spectroscopic parallax to be more reliable.  HD 36869 does feature extremely high levels of activity, abundant lithium, and rapid rotation consistent with the young ($\sim$30 Myr) age of Columba.  SEEDS does not detect any companion candidates within $7.\!\!''5$, $\sim$400 AU projected. 

{\it HIP 29067 (= GJ 9198)}---This K8 star is associated with the Castor moving group.  HIP 29067 shows modest Ca\,{\sc ii} HK activity, but was not detected in the {\it ROSAT} all-sky survey and has little lithium.  HS Psc, a $\sim$100 Myr-old mid-K star in our sample, shows much stronger activity and lithium absorption.  HIP 29067 is likely a much older star than its proposed Castor membership would imply.  SEEDS imaging does not detect any companion candidates within $7.\!\!''5$ ($\sim$180 AU projected).

{\it HIP 30030 (= HD 43989)}---This F9 star has been classified both as a member of TW Hydrae \citep{Zuckerman+Song_2004} and Columba \citep{Malo+Doyon+Lafreniere+etal_2013}.  We adopt the newer classification, which favors membership in Columba due to HIP 30030's Galactic position.  Both associations are young ($\sim$8 Myr for TW Hydrae, $\sim$30 Myr for Columba), and HIP 30030 has strong youth indicators.  SEEDS detects one highly significant companion candidate at a separation of $2.\!\!''57$; however, archival images from NICMOS reveal it as a background star.  

{\it HIP 32104 (= HD 48097)}---This A2 star is a reliable member of the Columba moving group.  Primarily part of the SEEDS high-mass sample, it is included here for completeness.  SEEDS images do not detect any companions.

{\it V429 Gem}---This K5 star is a reliable member of the AB Dor moving group, and shows strong youth indicators.  Radial velocity surveys have detected a 6.5 $M_J$ companion on a 7.8 day orbit \citep{Hernan-Obispo+Galvez-Ortiz+Anglada-Escude+etal_2010}.  However, V429 Gem's strong activity make radial velocity measurements difficult, and other authors have disputed the existence of a companion \citep{Figueira+Marmier+Bonfils+etal_2010}.  SEEDS images reveal a bright background star at a separation of $6.\!\!''97$ and a considerably fainter candidate at $({\rm E},\,{\rm N}) = (-1.\!\!''92,\,3.\!\!''19)$.  Follow-up observations revealed that this candidate is also a background object.  No other candidates were detected within $7''$ ($\sim$180 AU projected).  

{\it HIP 37288 (= GJ 281)}---This K7 star was originally proposed to be a member of the Local Association \citep{Montes+Lopez-Santiago+Galvez+etal_2001}, but was later reclassified in the Her-Lya association \citep{Lopez-Santiago+Montes+Crespo-Chacon+etal_2006}.  HIP 37288 shows only weak chromospheric activity and was not detected in the {\it ROSAT} all-sky survey; its lithium absorption lines are also weaker than the mid-K stars in our sample reliably associated with young, coeval moving groups.  SEEDS images reveal a companion candidate at $({\rm E},\,{\rm N}) = (-3.\!\!''95,\,-1.\!\!''64)$; however, archival images from Gemini/NIRI reveal it to be an unrelated background star.  

{\it HIP 39896 (= GJ 1108 A)}---This K7 star was originally proposed to be a member of the Local Association.  However, BANYAN indicates a possible membership in Columba, with $\sim$70\% probability neglecting $I$ and $J$ photometry.  HIP 39896 also has abundant youth indicators, including very rapid rotation and high chromospheric and coronal activity.  We consider it a possible Columba member and provisionally assign a 50\% membership probability.  SEEDS has revealed, for the first time, a close binary companion, with a separation of $0.\!\!''25$.  With an $H$-band contrast of only a factor of $\sim$6.4, the companion is likely to be an early M-dwarf.  HIP 39896 is also bound to a spectroscopic M2.8+M3.3 binary (GJ 1108 B) at a separation of $14''$ \citep{Lepine+Bongiorno_2007, Shkolnik+Hebb+Liu+etal_2010}, $\sim$300 AU projected, making HIP 39896 part of a hierarchical quadruple system.  

{\it HIP 40774}---This G5 star was proposed as a member of the IC 2391 supercluster.  Archival data in the literature, including a non-detection by {\it ROSAT}, and weak photospheric lithium absorption, cast further doubt on the star's youth.  \cite{Mishenina+Soubiran+Bienayme+etal_2008} report a lithium abundance $\log n({\rm Li}) = 1.6$ on the scale with ${\rm H} = 12$, which would be consistent with the values reported by \cite{Sestito+Randich_2005} for stars of similar $T_{\rm eff}$ in clusters of several Gyr age.  Taken together, these data suggest an age for HIP 40774 of $\gtrsim$1 Gyr.  SEEDS images detect a companion candidate at a separation of $({\rm E},\,{\rm N}) = (-0.\!\!''10,\,-4.\!\!''48)$; follow-up observations showed it to be an unrelated background object.  

{\it HIP 44526 (= HD 77825)}---This K2 star has been classified as a member of the Castor moving group.  It shows only modest levels of activity and lithium absorption, but does have a well-measured period.  SEEDS images show two bright companion candidates at a separation of $\sim$$7.\!\!''5$; follow-up imaging concluded showed that both were unrelated background stars.

{\it HIP 45383 (= HD 79555 = GJ 339)}---This K3 star has been classified as a member of the Castor moving group.  It does, however, have chromospheric measurements and X-ray activity consistent with a reasonably young age.  SEEDS images reveal the system to be a binary with an angular separation of $0.\!\!''74$ and an $H$-band flux ratio of $\sim$4.1: a K-dwarf and an M-dwarf with a projected separation of 13 AU.
SEEDS images detected a more distant companion candidate at a separation of $7''$; however, follow-up observations revealed it to be an unrelated background star.

{\it HIP 46843 (= HD 82443 = GJ 354.1)}---This K0 star was originally classified in the Local Association.  However, BANYAN indicates that it is a likely member of Columba, estimating a membership probability of just over 95\%.  HIP 46843 also has an extraordinary suite of secondary age indicators, including a rotation period.  The star's abundant lithium, rapid rotation, and high level of activity confirm its youth, and we estimate a 90\% probability of bona fide membership in Columba.  SEEDS images detect a companion candidate with a separation of $4''$; however, archival images from Gemini/NIRI confirm its status as an unrelated background object.

{\it HIP 50156 (= GJ 2079)}---This M0 star was recently proposed as a member of $\beta$ Pic \citep{Schlieder+Lepine+Simon_2012a}; however, the recent Bayesian analysis of \cite{Malo+Doyon+Lafreniere+etal_2013} finds a better match to the Columba moving group.  While the identity of its parent group remains ambiguous, the star is very active and unlikely to be a member of the field.  \cite{Malo+Doyon+Lafreniere+etal_2013} mention a surprisingly large scatter in the radial velocity measurements necessary to clarify membership and suggest that HIP 50156 may be a spectroscopic binary.  However, deep SEEDS images show no evidence for a stellar companion outside $\sim$$0.\!\!''02$, $\sim$0.5 AU.  Assuming that HIP 50156 is a spectroscopic binary, its two components must be very close, potentially accounting for the strong observed activity (and large scatter in reported $R'_{\rm HK}$ values), and the system may be tidally locked.  We tentatively consider it to be a member of Columba, although the similar ages of Columba and $\beta$ Pic make the distinction somewhat minor for our purposes.  SEEDS images detect no companion candidates within $7.\!\!''5$, $\sim$170 AU projected.  

{\it GJ 388}---This M4 star has been classified in Castor.  It shows significant activity and rapid rotation, but as an M star, these are difficult to use as indicators of youth.  SEEDS images do not reveal any companion candidates.  However, unfortunately, this data set features an exceptionally small amount of field rotation $2^\circ$.  The target passed almost directly overhead but was not successfully tracked until after it had passed zenith.  

{\it HIP 50660}---This K0 star has been proposed as a member of the IC 2391 supercluster.  HIP 50660 was not detected by {\it ROSAT} and has few other measurements in the literature.  SEEDS images detected a companion candidate with 5.9$\sigma$ significance at a separation of $4.\!\!''2$.  However, slightly deeper follow-up did not recover the point source, making it a likely statistical fluctuation.

{\it HIP 51317 (= GJ 393)}---This M2 star is a reliable member of AB Dor.  SEEDS images detect no companion candidates within $7.\!\!''5$ ($\sim$50 AU projected).  

{\it HIP 53020 (= GJ 402)}---This nearby M5 star has been classified in Her-Lya.  There is little additional data on the star, and its late spectral type makes any sort of dating extremely difficult.  SEEDS images detect no companion candidates within $8.\!\!''5$ ($\sim$60 AU projected). 

{\it HIP 53486 (= HD 94765 = GJ 3633)}---This K0 star has been classified as a member of the Castor moving group.  Secondary age indicators show moderate levels of chromospheric and coronal activity together with a modest rotation period, and probably indicate an older age.  SEEDS images detect no companion candidates within $7.\!\!''5$ ($\sim$130 AU projected).

{\it HD 95174}---This K2 star, together with its K5 binary companion, has recently been proposed as a member of $\beta$ Pic.  It was not detected by {\it ROSAT}, but was instead selected as a candidate member based on its strong UV emission, and confirmed as a likely member based on its Galactic motion \citep{Schlieder+Lepine+Simon_2012b}.  New spectroscopy, however, indicates an almost complete depletion of photospheric lithium, which is not expected for such a young K-dwarf.  Chromospheric activity measurements are too uncertain to strongly constrain the system's youth.  Further, as shown in Figure 1 of \cite{Schlieder+Lepine+Simon_2012b}, its UVW velocity would place it right at the edge of the $\beta$ Pic search area in all three velocity components, and it lies about 30 pc above the bulk of the bona fide $\beta$ Pic members.  With further age constraints from our spectroscopic follow-up, we consider HD 95174's classification in $\beta$ Pic to be highly doubtful.  SEEDS images reveal no companion candidates other than the known K5 secondary at a separation of $5''$.

{\it HIP 54155 (= HD 96064)}---This G8 star is a proposed member of the Local Association.  Though we do not infer an age from this, HIP 54155's secondary age indicators show high levels of activity and relatively abundant lithium.  SEEDS images show a bright background star also detected in archival Gemini/NIRI images, together with a marginal, 5.5$\sigma$ source at $({\rm E},\,{\rm N}) = (-3.\!\!''28,\,-3.\!\!''23)$, $1''$ from the background star.  In spite of their comparable sensitivity, the Gemini/NIRI images do not show this fainter source at either the same relative position or at the expected background position; it is almost certainly a statistical fluctuation. 

{\it TWA 2}---This M2 star is a young T Tauri object in the TW Hydrae moving group.  It is a binary, with its two components of similar brightness and separated by $0.\!\!''4$.  SEEDS observations were conducted in poor conditions, especially since the object's declination of $-30^\circ$ makes it relatively inaccessible from Subaru.  Less than 1 minute of integration time was obtained before the observation was abandoned. We have therefore omitted TWA 2 from our contrast tables.

{\it TYC 3825-716-1}---This K7 star was recently proposed as a member of the AB Dor moving group.  It is detected by {\it ROSAT} and shows strong UV emission consistent with youth \citep{Schlieder+Lepine+Simon_2012b}.  TYC 3825-716-1 lies on the outskirts of the AB Dor moving group in UVW velocity space, and $\sim$30 pc above most of the stars in Galactic Z distance.  As a result, BANYAN gives a negligible probability for AB Dor membership.  Our spectroscopy detects modest lithium absorption, but shows a chromospherically inactive star.  Lacking a parallax or more compelling secondary age indicators, we decline to assign the star any probability of membership in AB Dor.  SEEDS images do not detect any companions within $8''$ ($\sim$460 AU projected assuming the star's distance as inferred from kinematics and assuming AB Dor membership).

{\it HIP 59280 (= HD 105631 = GJ 3706)}---This K0 star has been classified in the IC 2391 supercluster.  Its secondary age indicators show only modest chromospheric and coronal activity and lithium absorption, measurements consistent with an age closer to 1 Gyr.  SEEDS images detect no companion candidates within $7.\!\!''5$ ($\sim$190 AU projected). 

{\it TYC 4943-192-1}---This M0 star was recently proposed to be a member of the AB Dor moving group \citep{Schlieder+Lepine+Simon_2010}.  \cite{Malo+Doyon+Lafreniere+etal_2013} confirmed it as an excellent candidate, but without a trigonometric parallax and with few secondary age indicators, a conclusive association is not yet possible.  We consider TYC 4943-192-1 to be a likely member of AB Dor, provisionally adopting the $\sim$80\% membership probability suggested by \cite{Malo+Doyon+Lafreniere+etal_2013}.  SEEDS images do not detect any companion candidates within $8.\!\!''5$ ($\sim$250 AU projected, assuming the distance inferred from kinematics and AB Dor membership).

{\it HIP 60661 (= GJ 466)}---This M0 star is a proposed member of the Local Association \citep{Montes+Lopez-Santiago+Fernandez-Figueroa+etal_2001}.  \cite{Lopez-Santiago+Montes+Crespo-Chacon+etal_2006} suggest membership in AB Dor, though HIP 60661 has the largest discrepancy in V and W with AB Dor's average kinematics of all of their proposed members.  \cite{Lopez-Santiago+Montes+Galvez-Ortiz+etal_2010} note a significantly discrepant radial velocity in their measurements of HIP 60661 and suggest that it may be a binary.  It has few secondary age indicators, and its weak chromospheric activity may indicate a much older age than that inferred for a young moving group.  We do not consider the star to be a likely member of AB Dor.  SEEDS images confirm HIP 60661's binary status, with a companion separated by $1.\!\!''9$ and an $H$-band contrast of $\sim$5.7 ($\sim$70 AU projected).  The companion detected by SEEDS is too far away, however, to account for the observed variation in radial velocity of several km\,s$^{-1}$ over a period of a few years.  There may be another, as-yet-undetected, companion lurking much closer to HIP 60661.  SEEDS images do not detect any other companion candidates within $7.\!\!''5$ ($\sim$280 AU projected). 

{\it HIP 63317 (= HD 112733)}---This G5 star is a proposed member of the Local Association.  HIP 63317 does show strong coronal and chromospheric activity and significant lithium, which provide some constraint on its age.  SEEDS images detect no companion candidates within $7.\!\!''5$ ($\sim$330 AU projected).  

{\it FH CVn}---This K7 star was recently proposed as a member of the AB Dor moving group \citep{Schlieder+Lepine+Simon_2012b}.  Its Galactic velocity is in excellent agreement with the bona fide members of the group, though FH CVn lies somewhat above the Galactic plane compared with the more established members.  The star also shows very strong X-ray activity and rapid rotation.  However, our spectroscopy indicates weak chromospheric activity and little photospheric lithium, giving some tension between different age indicators.  FH CVn does lie at the lithium depletion boundary at an age of $\sim$130 Myr \citep{Mentuch+Brandeker+vanKerwijk+etal_2008}, making the weakness of photospheric lithium absorption somewhat expected.  We provisionally consider it to be a moderately likely member of AB Dor, though additional follow-up is needed to clarify FH CVn's status.  Using only the available kinematic data, BANYAN estimates a 40\% probability of AB Dor membership, which we adopt for our analysis.  SEEDS images do not detect any companions within $7.\!\!''5$ ($\sim$350 AU assuming the distance inferred from moving group membership).

{\it HIP 66252 (= HD 118100 = GJ 517)}---This K5 star has been classified in the IC 2391 supercluster.  It has an extraordinary range of secondary age indicators, including vigorous activity, rapid rotation, and relatively abundant lithium, all of which point to youth.  SEEDS imaging detects no companion candidates within $7.\!\!5''$, $\sim$150 AU projected.  

{\it HIP 67412 (= HD 120352)}---This K0 star has been classified in the IC 2391 supercluster.  Its secondary age indicators show only modest chromospheric and coronal activity, while our spectroscopy reveals little photospheric lithium.  SEEDS images reveal no companion candidates within $7.\!\!''5$, $\sim$280 AU projected.  

{\it HIP 73996 (= HD 134083 = GJ 578)}---This F5 star was classified as a member of the UMa supercluster by \cite{Montes+Lopez-Santiago+Fernandez-Figueroa+etal_2001}.  However, more recent work disputes this classification.  \cite{Maldonado+Martinez-Arnaiz+Eiroa+etal_2010} list HIP 73996 as a probable nonmember, while their Table 8 claims that \cite{Lopez-Santiago+Montes+Galvez-Ortiz+etal_2010} list it as a probable member (the latter paper does not include the star at all).  The star is not especially active either chromospherically or coronally.  Our spectroscopy shows little photospheric lithium.  Regardless of its membership in the UMa supercluster, HIP 73996 is almost certainly not a member of the coeval UMa moving group.  SEEDS images revealed a 7$\sigma$ companion candidate at $({\rm E},\,{\rm N}) = (2.\!\!''76,\,-0.\!\!''33)$ with an $H$-band contrast of $2 \times 10^6$.  However, this candidate was near an image artifact and was not detected in follow-up observations.  No other candidates were detected within $7.\!\!''5$ ($\sim$280 AU projected).  

{\it HIP 78557 (= HD 143809)}---This G0 star is a proposed member of the Local Association.  At 80 pc, it is also the most distant target in our sample.  HIP 78577 has modest chromospheric, coronal activity, and reasonably abundant lithium.  However, its relatively early spectral type makes these age indicators somewhat less useful.  SEEDS detects a binary companion with a separation of $0.\!\!''57$ ($\sim$45 AU projected) and an $H$-band contrast of $\sim$190, which would make it a late M-dwarf.  

{\it HIP 82688 (= HD 152555)}---This late F/early G star is considered a reliable member of AB Dor, and its secondary age indicators confirm its youth.  SEEDS images detect two bright stars, one of which is HIP 82688's binary companion.  The companion has a separation of $3.\!\!''8$ ($\sim$85 AU projected) and an $H$-band contrast of $\sim$41, making it likely a mid-M dwarf.  SEEDS also detects a faint candidate at $({\rm E},\,{\rm N}) = (-1.\!\!''62,\,1.\!\!''82)$; however, follow-up observations revealed it to be a background star. 

{\it HIP 83494 (= 154431)}---This A5 star was proposed as a member of Tuc-Hor by \cite{Zuckerman+Rhee+Song+etal_2011}.  While it shows evidence of a debris disk, and is likely young, \cite{Malo+Doyon+Lafreniere+etal_2013} find a very poor match to Tuc-Hor, and classify HIP 83494 as a field star (nonmember of any of their studied associations) with high confidence.  We decline to place a nonzero probability on Tuc-Hor membership.  HIP 83494 is a member of the SEEDS high-mass sample, and is included here for completeness.  The secondary age indicators that we use for our other targets are of little use for such a high-mass star.  SEEDS images detected no companions.

{\it HIP 87579 (= GJ 697)}---This K0 star is a proposed member of the Castor moving group.  Secondary age indicators show only modest coronal and chromospheric activity.  SEEDS images detect many companion candidates within $8''$ ($\sim$200 AU projected); however, follow-up observations reveal them all to be background stars.  

{\it HIP 87768 (= GJ 698)}---This K5 star is a proposed member of the Local Association.  SEEDS observations do not detect any companion candidates within $7.\!\!''5$ ($\sim$190 AU projected).

{\it HIP 91043 (= HD 171488)}---This G2 star is a proposed member of the Local Association.  Its secondary age indicators show an exceptionally active star with abundant lithium and rapid rotation.  The isochrone analysis shows a strong discrepancy at first glance, due to two strong peaks in the likelihood function, at $\sim$20 Myr and $\sim$10 Gyr.  The latter age is certainly incompatible with the secondary age indicators.  The isochrones may therefore indicate that our age estimates are overly conservative.  SEEDS observations reveal many companion candidates; however, this is expected for a source at low Galactic latitude, with $(l,\,b) = (48^\circ,\,12^\circ)$.  Follow-up observations confirmed these sources to be background objects; several brighter sources were seen in Keck imaging and listed in \cite{Metchev+Hillenbrand_2009}.

{\it HIP 93580 (= HD 177178)}---This A4 star was proposed as a member of AB Dor by \cite{Zuckerman+Rhee+Song+etal_2011}.  It is part of the SEEDS high-mass sample, and included here for completeness.  \cite{Malo+Doyon+Lafreniere+etal_2013} also favor membership in AB Dor, but with a somewhat low probability of 80\% neglecting photometry, due to a discrepant U velocity.  The isochrone analysis further calls this moving group assignment into question.  The likelihood function shows two peaks: one from $\sim$5--20 Myr, and a second, broad peak centered at $\sim$500 Myr.  We provisionally adopt a lower probability of 30\% for group membership.  Due to HIP 93580's high mass, it never develops a large outer convective zone, and secondary age indicators are of little value.  HIP 93580 lies just $2^\circ$ from the Galactic plane and has many companion candidates awaiting follow-up observations.

{\it BD$+$05 4576}---This K7 star has recently been proposed as a member of AB Dor \citep{Schlieder+Lepine+Simon_2010} on the basis of kinematics and its X-ray flux as measured by {\it ROSAT}.  However, there is no trigonometric parallax, and our spectroscopy finds little photospheric lithium.  The latter is not too surprising, as the star's spectral type places it right at the lithium depletion boundary for an age of $\sim$130 Myr.  Given the paucity of data, we adopt the 80\% membership probability as estimated by BANYAN based only on the available kinematics.  SEEDS images do not detect any companion candidates within $8''$ ($\sim$300 AU assuming the distance inferred from kinematics and group membership).

{\it HIP 102409 (= HD 197481 = GJ 803 = AU Mic)}---This M1 star hosts a well-known debris disk, and is reliably identified with the $\beta$ Pic moving group.  The debris disk appears nearly edge-on and extends out to 200 AU in radius \citep{Kalas+Liu+Matthews_2004}, making AU Mic an excellent target for high-contrast observations.  However, its declination of $-31^\circ$ makes it difficult to observe from Subaru; it is the most southerly target in the entire moving group sample.  SEEDS images do not detect any companions within $3.\!\!''2$, $\sim$30 AU projected.

{\it HD 201919}---This K6 star is a likely member of the AB Dor moving group, and its secondary age indicators confirm its youth.  SEEDS images detect only a bright companion candidate just over $7''$ away, a projected separation of nearly 300 AU at HD 201919's distance inferred from its kinematics assuming moving group membership.  

{\it HIP 107350 (= HD 206860 = GJ 9751)}---This G0 star is a proposed member of the Her-Lya Association.  HIP 107350 does, however, have an exceptional set of measurements in the literature, including a rotation period and multi-decade Mt.~Wilson chromospheric data.  These secondary age indicators point to a relatively young system.  SEEDS observations detect a single companion candidate around HIP 107350, but archival images from Gemini/NIRI reveal it to be an unrelated background star.  

{\it TYC 2211-1309-1}---This K7 star was recently proposed as a likely member of $\beta$ Pic based on its kinematics and strong secondary youth indicators.  Indeed, this star is the fastest known rotator in our sample, with a period less than 0.5 days independently measured by \cite{Norton+Wheatley+West+etal_2007} and \cite{Messina+Desidera+Tutatto+etal_2010}.  However, our new spectroscopic measurements introduce some tension with the strong X-ray activity and rapid rotation, finding little evidence of photospheric lithium.  This may point to an unusual accretion history or recent merger responsible for both the high angular momentum and relative lack of lithium; TYC 2211-1309-1 certainly deserves more study.  We consider it to be a likely, albeit far from certain, member of $\beta$ Pic, and provisionally assign a 50\% membership probability.  SEEDS images were taken under very poor observing conditions and were made in the $K$-band to enable even a basic AO correction.  The only companion candidate detected was also visible in archival NACO images, which showed it to be an unrelated background star.

{\it HIP 111449 (= HD 213845 = GJ 863.2)}---This F7 star is a proposed member of the Her-Lya Association.  It has limited activity measurements from the literature, is relatively inactive in X-rays, and lacks significant photospheric lithium.  SEEDS detects a stellar companion at a separation of $6.\!\!''1$ ($\sim$140 AU projected); this object was previously reported in \cite{Lafreniere+Doyon+Marois+etal_2007}.  

{\it HIP 114066 (= GJ 9809)}---This M0 star is a reliable member of the AB Dor moving group, and shows relatively rapid rotation and vigorous X-ray activity.  HIP 114066 lies nearly in the Galactic plane; as a result, it has an extremely high density of spurious background stars.  Using archival Gemini/NIRI data, we have confirmed that all of these candidates are unrelated background stars.  

{\it HIP 115162}---This star is a reliable member of the AB Dor moving group.  Its secondary age indicators show strong signs of youth, including abundant photospheric lithium and relative strong chromospheric and coronal activity.  HIP 115162 has had some controversy over its spectral type, with \cite{Schlieder+Lepine+Simon_2010} listing it as G0V, while \cite{Zuckerman+Song_2004} list G4 and \cite{Ofek_2008} fit G8V to an SED template.  We used the known spectral-class/temperature dependent line ratio of Fe\,{\sc ii} 6432.65\AA/Fe\,{\sc i} 6430.85\AA \citep{Strassmeier+Fekel_1990,Montes+Martin_1998} to better constrain the spectral classification of HIP 115162.  The observed line ratio for 
HIP 115162, $\sim$0.22, was more consistent with that observed in K0V stars ($\sim$0.2; \citealt{Montes+Martin_1998}) than in G0V ($\sim$0.5; 
\citealt{Montes+Martin_1998}) or G5V ($\sim$0.4; \citealt{Montes+Martin_1998}) stars.  This supports a late G spectral type for HIP 115162, and we adopt the \cite{Ofek_2008} G8V classification.  SEEDS images detect no companion candidates within $7''$, or 350 AU projected. 

{\it BD$-$13 6424}---This M0 star is a reliable member of the $\beta$ Pic moving group.  Its secondary age indicators show rapid rotation, abundant lithium, and strong X-ray activity.  SEEDS images detect no companion candidates within $7.\!\!''5$, or $\sim$200 AU projected.

{\it HIP 116805 (= HD 222439 = $\kappa$ And)}---{\sc banyan} gives a very high probability, 95\%, of Columba membership, as was asserted in its companion's discovery paper \citep{Carson+Thalmann+Janson+etal_2013}.  However, our isochrone analysis casts doubt on this classification, with a strong peak in the likelihood function at $\sim$200 Myr.  Other authors have recently re-analyzed HIP 116805 and also find evidence for an older age and possible non-membership in Columba \citep{Bonnefoy+Currie+Marleau+etal_2013, Hinkley+Pueyo+Faherty+etal_2013}. We note however, that the rapid rotation and unknown inclination angle of the star may make isochronal age determination unreliable; if the star is viewed close to pole on, it could be as young as Columba.  In this work, we provisionally assign the star a 30\% probability of Columba membership.  HIP 116805 is primarily part of the SEEDS high-mass sample, and is included here for completeness.  HIP 116805 hosts a substellar companion, $\kappa$ And b, recently discovered by SEEDS \citep{Carson+Thalmann+Janson+etal_2013}.  $\kappa$ And b has a mass of $\sim$13--50 $M_J$, depending on the assumed system age \citep{Carson+Thalmann+Janson+etal_2013, Bonnefoy+Currie+Marleau+etal_2013, Hinkley+Pueyo+Faherty+etal_2013}, and lies at a separation of $1.\!\!''06$, or 55 AU projected, from its host star.

\begin{deluxetable*}{lcccccccr}
\tablewidth{0pt}
\vspace{-0.3truein}
\tablecaption{The SEEDS Moving Group Observing Log}
\tablehead{
    \colhead{Name} &
    \colhead{$\alpha$ (J2000)} &
    \colhead{$\delta$ (J2000)} &
    \colhead{$N_{\rm exp}$} &
    \colhead{$t_{\rm tot}$} &
    \colhead{Rot} &
    \colhead{Mean} &
    \colhead{Date} \\     
    \colhead{} &
    \colhead{(h m s)} &
    \colhead{($^\circ$ $'$ $''$)} &
    \colhead{} &
    \colhead{(min)} &
    \colhead{($^\circ$)} &
    \colhead{Airmass} &
    \colhead{y-m-d}    
}
    
\startdata 

HIP 544 & 00 06 36.8 & $+$29 01 17 & 325 & 16.3 & 76 & 1.02 & 2010-12-01 \\ 
HIP 1134 & 00 14 10.3 & $-$07 11 57 & 151 & 37.8 & 29 & 1.14 & 2011-08-02 \\ 
FK Psc & 00 23 34.7 & $+$20 14 29 & 98 & 24.5 & 172 & 1.00 & 2011-09-03 \\ 
HIP 3589 & 00 45 50.9 & $+$54 58 40 & 120 & 30.0 & 22 & 1.24 & 2011-12-30 \\ 
-------- & 00 45 50.9 & $+$54 58 40 & 138 & 46.0 & 25 & 1.25 & 2012-09-12 \\ 
HIP 4979 & 01 03 49.0 & $+$01 22 01 & 47 & 22.9 & 23 & 1.06 & 2009-11-02 \\ 
-------- & 01 03 49.0 & $+$01 22 01 & 80 & 13.3 & 13 & 1.06 & 2012-09-13 \\ 
HIP 6869 & 01 28 24.4 & $+$17 04 45 & 59 & 13.7 & 103 & 1.00 & 2009-11-02 \\ 
HS Psc & 01 37 23.2 & $+$26 57 12 & 258 & 43.0 & 87 & 1.01 & 2012-09-14 \\ 
HIP 10679 & 02 17 24.7 & $+$28 44 30 & 111 & 27.8 & 67 & 1.01 & 2011-12-24 \\ 
BD+30 397B & 02 27 28.0 & $+$30 58 41 & 116 & 38.7 & 68 & 1.02 & 2011-12-31 \\ 
HIP 11437 & 02 27 29.3 & $+$30 58 25 & 116 & 38.7 & 68 & 1.02 & 2011-12-30 \\ 
-------- & 02 27 29.3 & $+$30 58 25 & 129 & 32.3 & 60 & 1.03 & 2011-12-31 \\
HIP 12545 & 02 41 25.9 & $+$05 59 18 & 135 & 31.3 & 39 & 1.05 & 2009-12-24 \\ 
HIP 12638 & 02 42 21.3 & $+$38 37 07 & 120 & 40.0 & 30 & 1.10 & 2011-09-06 \\ 
HIP 12925 & 02 46 14.6 & $+$05 35 33 & 120 & 30.0 & 48 & 1.03 & 2012-01-01 \\ 
HIP 17248 & 03 41 37.3 & $+$55 13 07 & 81 & 40.5 & 24 & 1.25 & 2012-11-07 \\ 
HIP 23362 & 05 01 25.6 & $-$20 03 07 & 55 & 27.5 & 13 & 1.33 & 2012-11-07 \\
HIP 25486 & 05 27 04.8 & $-$11 54 03 & 120 & 11.1 & 12 & 1.18 & 2010-01-24 \\ 
HD 36869 & 05 34 09.2 & $-$15 17 03 & 81 & 40.5 & 26 & 1.24 & 2012-11-06 \\ 
HIP 29067 & 06 07 55.2 & $+$67 58 37 & 162 & 40.5 & 20 & 1.75 & 2012-04-11 \\ 
HIP 30030 & 06 19 08.1 & $-$03 26 20 & 87 & 29.0 & 15 & 1.18 & 2011-03-25 \\ 
HIP 32104 & 06 42 24.3 & $+$17 38 43 & 135 & 11.3 & 39 & 1.00 & 2011-12-25 \\
V429 Gem & 07 23 43.6 & $+$20 24 59 & 117 & 27.2 & 168 & 1.00 & 2010-01-23 \\ 
HIP 37288 & 07 39 23.0 & $+$02 11 01 & 124 & 31.0 & 40 & 1.06 & 2011-01-30 \\ 
HIP 39896 & 08 08 56.4 & $+$32 49 11 & 95 & 23.8 & 48 & 1.04 & 2011-12-25 \\ 
HIP 40774 & 08 19 19.1 & $+$01 20 20 & 121 & 28.1 & 27 & 1.08 & 2009-12-25 \\
-------- & 08 19 19.1 & $+$01 20 20 & 87 & 21.8 & 43 & 1.06 & 2011-01-28 \\ 
HIP 44526 & 09 04 20.7 & $-$15 54 51 & 37 & 12.3 & 7 & 1.35 & 2011-01-30 \\ 
-------- & 09 04 20.7 & $-$15 54 51 & 92 & 30.7 & 19 & 1.27 & 2012-01-01 \\
HIP 45383 & 09 14 53.7 & $+$04 26 34 & 90 & 30.0 & 47 & 1.04 & 2011-03-26 \\ 
HIP 46843 & 09 32 43.8 & $+$26 59 19 & 106 & 35.3 & 99 & 1.01 & 2011-01-28 \\ 
HIP 50156 & 10 14 19.2 & $+$21 04 30 & 137 & 34.3 & 166 & 1.00 & 2011-12-24 \\ 
GJ 388 & 10 19 36.3 & $+$19 52 12 & 105 & 26.3 & 2 & 1.06 & 2012-05-16 \\ 
HIP 50660 & 10 20 45.9 & $+$32 23 54 & 104 & 24.1 & 48 & 1.03 & 2009-12-23 \\ 
HIP 51317 & 10 28 55.6 & $+$00 50 28 & 118 & 19.7 & 29 & 1.08 & 2011-01-28 \\ 
HIP 53020 & 10 50 52.0 & $+$06 48 29 & 155 & 51.7 & 32 & 1.08 & 2011-01-29 \\ 
-------- & 10 50 52.0 & $+$06 48 29 & 95 & 23.8 & 36 & 1.03 & 2011-05-25 & \\
HIP 53486 & 10 56 30.8 & $+$07 23 19 & 223 & 20.7 & 53 & 1.03 & 2010-01-25 \\ 
HD 95174 & 10 59 38.3 & $+$25 26 15 & 112 & 28.0 & 66 & 1.01 & 2012-05-11 \\ 
HIP 54155 & 11 04 41.5 & $-$04 13 16 & 136 & 34.0 & 25 & 1.13 & 2011-05-26 \\ 

TYC 3825-716-1 & 11 20 50.5 & $+$54 10 09 & 101 & 33.7 & 22 & 1.22 & 2012-02-27 \\ 
-------- & 11 20 50.5 & $+$54 10 09 & 171 & 42.8 & 33 & 1.24 & 2011-12-26 \\
HIP 59280 & 12 09 37.3 & $+$40 15 07 & 208 & 33.8 & 43 & 1.08 & 2009-12-23 \\ 

TYC 4943-192-1 & 12 15 18.4 & $-$02 37 28 & 51 & 25.5 & 14 & 1.13 & 2011-02-01 \\ 
HIP 60661 & 12 25 58.6 & $+$08 03 44 & 66 & 15.3 & 29 & 1.02 & 2010-01-23 \\ 
-------- & 12 25 58.6 & $+$08 03 44 & 71 & 23.7 & 30 & 1.03 & 2011-05-21 \\
HIP 63317 & 12 58 32.0 & $+$38 16 44 & 131 & 32.8 & 42 & 1.07 & 2012-05-14 \\ 
FH CVn & 13 27 12.1 & $+$45 58 26 & 115 & 38.3 & 31 & 1.12 & 2012-02-26 \\ 
HIP 66252 & 13 34 43.2 & $-$08 20 31 & 112 & 37.3 & 26 & 1.14 & 2011-05-26 \\ 
-------- & 13 34 43.2 & $-$08 20 31 & 95 & 31.7 & 24 & 1.14 & 2012-05-12 \\ 
HIP 67412 & 13 48 58.2 & $-$01 35 35 & 157 & 36.4 & 38 & 1.08 & 2010-01-24 \\ 
HIP 73996 & 15 07 18.1 & $+$24 52 09 & 174 & 14.5 & 103 & 1.01 & 2011-03-26 \\ 
-------- & 15 07 18.1 & $+$24 52 09 & 300 & 25.0 & 106 & 1.01 & 2013-02-26 \\ 
-------- & 15 07 18.1 & $+$24 52 09 & 233 & 38.8 & 106 & 1.01 & 2013-02-27 \\ 
HIP 78557 & 16 02 22.4 & $+$03 39 07 & 213 & 35.5 & 41 & 1.05 & 2012-07-08 \\ 
-------- & 16 02 22.4 & $+$03 39 07 & 30 & 5.0 & 4 & 1.19 & 2013-05-20 \\
HIP 82688 & 16 54 08.1 & $-$04 20 25 & 140 & 35.0 & 33 & 1.10 & 2011-05-24 \\ 
-------- & 16 54 08.1 & $-$04 20 25 & 161 & 53.7 & 39 & 1.14 & 2012-04-11 \\
HIP 83494 & 17 03 53.6 & $+$34 47 25 & 36 & 18.0 & 15 & 1.10 & 2012-02-26 \\
HIP 87579 & 17 53 29.9 & $+$21 19 31 & 142 & 35.5 & 139 & 1.01 & 2011-05-22 \\ 
-------- & 17 53 29.9 & $+$21 19 31 & 210 & 52.5 & 166 & 1.01 & 2012-05-13 \\
HIP 87768 & 17 55 44.9 & $+$18 30 01 & 192 & 32.0 & 38 & 1.01 & 2012-07-07 \\ 
HIP 91043 & 18 34 20.1 & $+$18 41 24 & 180 & 30.0 & 132 & 1.01 & 2012-07-10 \\ 
-------- & 18 34 20.1 & $+$18 41 24 & 108 & 21.6 & 59 & 1.01 & 2013-05-18 \\
HIP 93580 & 19 03 32.3 & $+$01 49 08 & 150 & 25.0 & 22 & 1.06 & 2012-07-11 \\
BD+05 4576 & 20 39 54.6 & $+$06 20 12 & 102 & 34.0 & 32 & 1.05 & 2011-05-23 \\ 
-------- & 20 39 54.6 & $+$06 20 12 & 24 & 6.0 & 15 & 1.04 & 2012-09-12 & \\
HIP 102409 & 20 45 09.5 & $-$31 20 27 & 53 & 25.8 & 12 & 1.67 & 2009-11-01 \\ 
HD 201919 & 21 13 05.3 & $-$17 29 13 & 94 & 47.0 & 26 & 1.29 & 2012-11-07 \\ 
HIP 107350 & 21 44 31.3 & $+$14 46 19 & 137 & 17.1 & 72 & 1.01 & 2011-08-03 \\ 
TYC 2211-1309-1 & 22 00 41.6 & $+$27 15 14 & 138 & 34.5 & 82 & 1.01 & 2011-09-04 \\ 
HIP 111449 & 22 34 41.6 & $-$20 42 30 & 620 & 25.8 & 26 & 1.33 & 2012-11-06 \\ 
HIP 114066 & 23 06 04.8 & $+$63 55 34 & 105 & 52.5 & 22 & 1.40 & 2012-11-05 \\ 
HIP 115162 & 23 19 39.6 & $+$42 15 10 & 150 & 50.0 & 37 & 1.08 & 2012-09-13 \\ 
BD$-$13 6424 & 23 32 30.9 & $-$12 15 51 & 123 & 41.0 & 24 & 1.21 & 2011-08-03 \\ 
HIP 116805 & 23 40 24.5 & $+$44 20 02 & 246 & 20.5 & 14 & 1.18 & 2012-01-01 \\
-------- & 23 40 24.5 & $+$44 20 02 & 201 & 26.8 & 26 & 1.10 & 2012-07-08 
\enddata
\label{tab:observing_log}
\end{deluxetable*}

\begin{deluxetable*}{lccccccccr}
\tablewidth{0pt}
\tablecaption{SEEDS Moving Group $5.5\sigma$ Contrast Limits}
\tablehead{
    \colhead{Name} &
    \colhead{H} &
    \multicolumn{6}{c}{5.5$\sigma$ Contrast (mag)} \\
    \colhead{} &
    \colhead{(mag)} &
    \colhead{$0.\!\!''25$} &
    \colhead{$0.\!\!''5$} &
    \colhead{$0.\!\!''75$} &
    \colhead{$1''$} &
    \colhead{$1.\!\!''5$} &
    \colhead{$2''$} &
    \colhead{$3''$} &
    \colhead{$5''$}
}
\startdata    
HIP 544 & $3.95 \pm 0.02$ & \ldots & 9.1 & 11.2 & 12.6 & 14.1 & 14.6 & 14.8 & 14.8  \\
HIP 1134 & $2.80 \pm 0.05$ & \ldots & 10.1 & 12.1 & 13.4 & 14.8 & 15.3 & 15.5 & 15.6  \\
FK Psc & $3.62 \pm 0.06$ & 5.9 & 7.8 & 9.0 & 9.8 & 8.7 & 10.1 & 11.7 & 12.0  \\
HIP 3589 & $2.80 \pm 0.10$ & \ldots & 9.0 & 11.1 & 12.5 & 13.5 & 12.8 & 14.1 & 14.7  \\
HIP 4979 & $1.62 \pm 0.05$ & \ldots & 10.3 & 11.8 & 13.7 & 14.9 & 15.5 & 15.8 & 15.8  \\
HIP 6869 & $1.75 \pm 0.06$ & \ldots & 8.2 & 11.7 & 13.2 & 14.5 & 14.8 & 14.8 & 14.5  \\
HS Psc & $4.85$ & 7.0 & 9.0 & 10.6 & 11.6 & 12.4 & 12.6 & 12.7 & 12.6  \\
HIP 10679 & $4.18 \pm 0.35$ & \ldots & 8.8 & 10.6 & 11.9 & 13.2 & 13.7 & 13.9 & 13.8  \\
BD+30 397B & $5.13 \pm 0.20$ & 7.7 & 9.6 & 11.5 & 12.5 & 13.2 & 13.4 & 13.4 & 13.3  \\
HIP 11437 & $4.23 \pm 0.20$ & 7.5 & 9.6 & 11.6 & 12.7 & 13.7 & 13.9 & 13.9 & 13.9  \\
HIP 12545 & $4.11 \pm 0.14$ & 7.8 & 10.0 & 11.8 & 13.0 & 13.7 & 14.0 & 14.1 & 14.1  \\
HIP 12638 & $3.81 \pm 0.23$ & \ldots & 9.4 & 11.4 & 12.6 & 13.7 & 14.1 & 14.3 & 14.5  \\
HIP 12925 & $2.96 \pm 0.12$ & 6.2 & 8.2 & 10.2 & 11.4 & 11.5 & 10.8 & 13.5 & 13.7  \\
HIP 17248 & $4.92 \pm 0.17$ & \ldots & 9.8 & 11.5 & 12.8 & 13.6 & 13.8 & 13.9 & 13.9  \\
HIP 23362 & $1.10 \pm 0.03$ & \ldots & 8.8 & 10.5 & 12.3 & 14.2 & 15.3 & 15.9 & 16.1  \\
HIP 25486 & $2.93 \pm 0.03$ & \ldots & 9.1 & 10.8 & 12.5 & 13.9 & 14.4 & 14.6 & 14.7  \\
HD 36869 & $4.26 \pm 0.54$ & \ldots & 8.8 & 10.5 & 12.0 & 13.4 & 13.9 & 14.1 & 14.2  \\
HIP 29067 & $4.86 \pm 0.10$ & \ldots & 9.1 & 11.2 & 12.6 & 13.9 & 14.2 & 14.4 & 14.5  \\
HIP 30030 & $3.13 \pm 0.09$ & \ldots & 7.8 & 9.6 & 11.2 & 12.7 & 13.3 & 13.5 & 13.7  \\
HIP 32104 & $1.87 \pm 0.06$ & \ldots & 7.0 & 8.2 & 9.6 & 11.1 & 12.0 & 12.5 & 12.7  \\
V429 Gem & $4.97 \pm 0.34$ & 7.0 & 9.8 & 11.4 & 12.5 & 13.2 & 13.4 & 13.5 & 13.2  \\
HIP 37288 & $5.27 \pm 0.04$ & \ldots & 9.0 & 10.7 & 12.2 & 13.8 & 14.7 & 15.0 & 15.1  \\
HIP 39896 & $5.00 \pm 0.15$ & 4.2 & 8.0 & 9.6 & 10.9 & 12.0 & 12.5 & 12.6 & 12.6  \\
HIP 40774 & $4.42 \pm 0.07$ & \ldots & 9.9 & 11.9 & 13.4 & 14.6 & 15.1 & 15.2 & 15.2  \\
HIP 44526 & $4.28 \pm 0.05$ & \ldots & 8.1 & 10.2 & 11.7 & 13.2 & 13.9 & 14.2 & 14.3  \\
HIP 45383 & $4.12 \pm 0.06$ & \ldots & 6.7 & 6.7 & 9.6 & 13.3 & 14.6 & 15.3 & 15.4  \\
HIP 46843 & $3.99 \pm 0.02$ & \ldots & 10.3 & 12.3 & 13.9 & 15.3 & 15.8 & 16.0 & 16.0  \\
HIP 50156 & $4.63 \pm 0.09$ & \ldots & 9.3 & 10.9 & 12.1 & 13.5 & 13.9 & 14.1 & 13.9  \\
GJ 388 & $6.48 \pm 0.05$ & \ldots & \ldots & \ldots & \ldots & 10.9 & 12.4 & 14.0 & 15.1  \\
HIP 50660 & $4.01 \pm 0.13$ & \ldots & 9.7 & 11.6 & 12.9 & 13.6 & 13.8 & 13.8 & 13.7  \\
HIP 51317 & $6.35 \pm 0.03$ & \ldots & 8.7 & 10.7 & 12.6 & 14.0 & 14.6 & 14.9 & 14.9  \\
HIP 53020 & $7.55 \pm 0.06$ & \ldots & 10.6 & 12.1 & 13.6 & 14.6 & 14.9 & 15.1 & 15.1  \\
HIP 53486 & $4.16 \pm 0.04$ & \ldots & 9.6 & 11.6 & 13.1 & 14.4 & 15.0 & 15.2 & 15.2  \\
HD 95174 & $4.19 \pm 0.19$ & \ldots & 9.8 & 11.7 & 13.1 & 14.2 & 14.6 & 14.5 & 11.7  \\
HIP 54155 & $3.80 \pm 0.06$ & \ldots & 8.6 & 10.3 & 11.7 & 13.4 & 14.1 & 14.4 & 14.6  \\
TYC 3825-716-1 & $4.88 \pm 0.21$ & \ldots & 9.1 & 10.6 & 11.5 & 12.1 & 12.1 & 12.2 & 12.2  \\
HIP 59280 & $3.75 \pm 0.04$ & \ldots & 9.8 & 11.8 & 13.3 & 14.6 & 15.0 & 15.2 & 15.3  \\
TYC 4943-192-1 & $5.60 \pm 0.19$ & \ldots & 8.7 & 10.3 & 11.3 & 12.2 & 12.6 & 12.7 & 12.7  \\
HIP 60661 & $4.45 \pm 0.19$ & \ldots & 9.0 & 10.4 & 11.3 & 10.5 & 9.5 & 12.2 & 12.2  \\
HIP 63317 & $3.72 \pm 0.13$ & 7.1 & 9.4 & 11.4 & 12.6 & 13.6 & 14.0 & 14.2 & 14.2  \\
FH CVn & $4.89 \pm 0.20$ & \ldots & 9.7 & 11.1 & 12.2 & 12.8 & 13.0 & 13.1 & 13.1  \\
HIP 66252 & $4.78 \pm 0.03$ & \ldots & 9.7 & 11.4 & 12.8 & 14.1 & 14.7 & 15.1 & 15.2  \\
HIP 67412 & $4.01 \pm 0.10$ & 7.7 & 10.5 & 12.3 & 13.4 & 14.1 & 14.4 & 14.4 & 14.4  \\
HIP 73996 & $2.55 \pm 0.01$ & \ldots & 9.8 & 11.6 & 12.9 & 14.4 & 15.3 & 15.8 & 16.0  \\
HIP 78557 & $2.95 \pm 0.26$ & 6.8 & 8.4 & 10.5 & 11.6 & 12.5 & 12.9 & 12.8 & 12.9  \\
HIP 82688 & $3.13 \pm 0.09$ & \ldots & 11.1 & 12.6 & 13.9 & 14.9 & 15.2 & 15.0 & 15.3  \\
HIP 83494 & $1.98 \pm 0.04$ & \ldots & 8.4 & 10.8 & 12.1 & 13.7 & 14.6 & 15.1 & 15.2  \\
HIP 87579 & $4.36 \pm 0.05$ & \ldots & 10.5 & 12.1 & 13.5 & 14.5 & 14.9 & 15.0 & 15.1  \\
HIP 87768 & $4.43 \pm 0.11$ & 6.3 & 8.0 & 9.8 & 11.1 & 12.5 & 13.3 & 13.7 & 14.0  \\
HIP 91043 & $3.00 \pm 0.05$ & \ldots & 9.1 & 10.7 & 12.0 & 13.5 & 14.0 & 14.0 & 15.1  \\
HIP 93580 & $1.66 \pm 0.04$ & \ldots & 9.2 & 11.0 & 12.4 & 13.7 & 14.4 & 14.5 & 14.6  \\
BD+05 4576 & $4.42$ & \ldots & 10.0 & 11.5 & 12.8 & 13.8 & 14.0 & 14.2 & 14.2  \\
HIP 102409 & $4.85 \pm 0.02$ & \ldots & 7.8 & 10.3 & 11.8 & 13.4 & 14.0 & 14.7 & \ldots  \\
HD 201919 & $4.79$ & \ldots & 9.1 & 11.0 & 12.3 & 13.2 & 13.5 & 13.7 & 13.7  \\
HIP 107350 & $3.34 \pm 0.01$ & \ldots & 10.0 & 11.9 & 13.5 & 14.8 & 15.4 & 15.5 & 15.5  \\
TYC 2211-1309-1 & $4.66 \pm 0.08$ & 6.3 & 8.2 & 9.4 & 10.6 & 11.1 & 11.3 & 11.3 & 11.1  \\
HIP 111449 & $2.49 \pm 0.01$ & \ldots & 8.9 & 10.8 & 12.3 & 14.2 & 15.0 & 15.6 & 15.7  \\
HIP 114066 & $5.22 \pm 0.09$ & \ldots & 9.8 & 11.6 & 12.9 & 14.0 & 14.4 & 14.6 & 14.6  \\
HIP 115162 & $3.78 \pm 0.13$ & 7.1 & 9.1 & 10.9 & 12.0 & 13.1 & 13.6 & 13.8 & 13.9  \\
BD$-$13 6424 & $4.59 \pm 0.03$ & \ldots & 9.3 & 10.9 & 12.5 & 13.8 & 14.3 & 14.6 & 14.7  \\
HIP 116805 & $1.04 \pm 0.02$ & \ldots & 9.4 & 10.8 & 12.2 & 14.1 & 15.1 & 15.8 & 15.9  
\enddata

\label{tab:contrast_curves}
\end{deluxetable*}

\begin{figure}
\includegraphics[width=\linewidth]{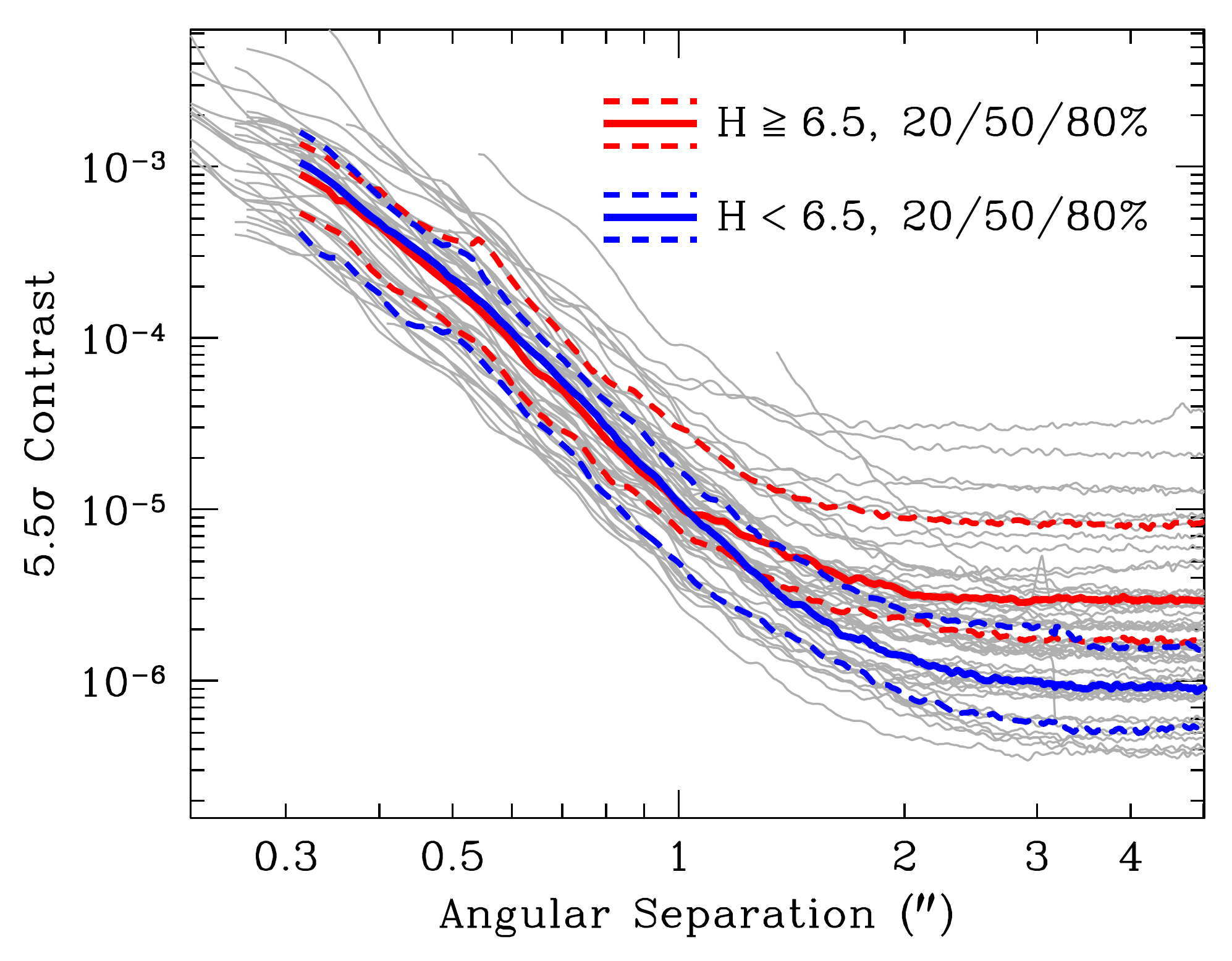}
\caption{The contrast curves for the SEEDS moving groups sample; FK Psc, HIP 3589, HIP 6869, HIP 12925, HIP 45383, HD 95174, HIP 60661, HIP 82688, and HIP 91043 show strong artifacts from bright neighbors and have been omitted.  At separations of $\lesssim$$1''$, the contrast limits depend on field rotation and observing conditions.  Several arcseconds from the star, SEEDS observations are read noise limited, and the magnitude limits depend on AO performance, total integration time, and integration per frame.  Fainter targets have less contrast, but fainter limiting magnitudes, at separations $\gtrsim2''$.}
\label{fig:contrast_curves}
\end{figure}

\begin{figure}
\includegraphics[width=\linewidth]{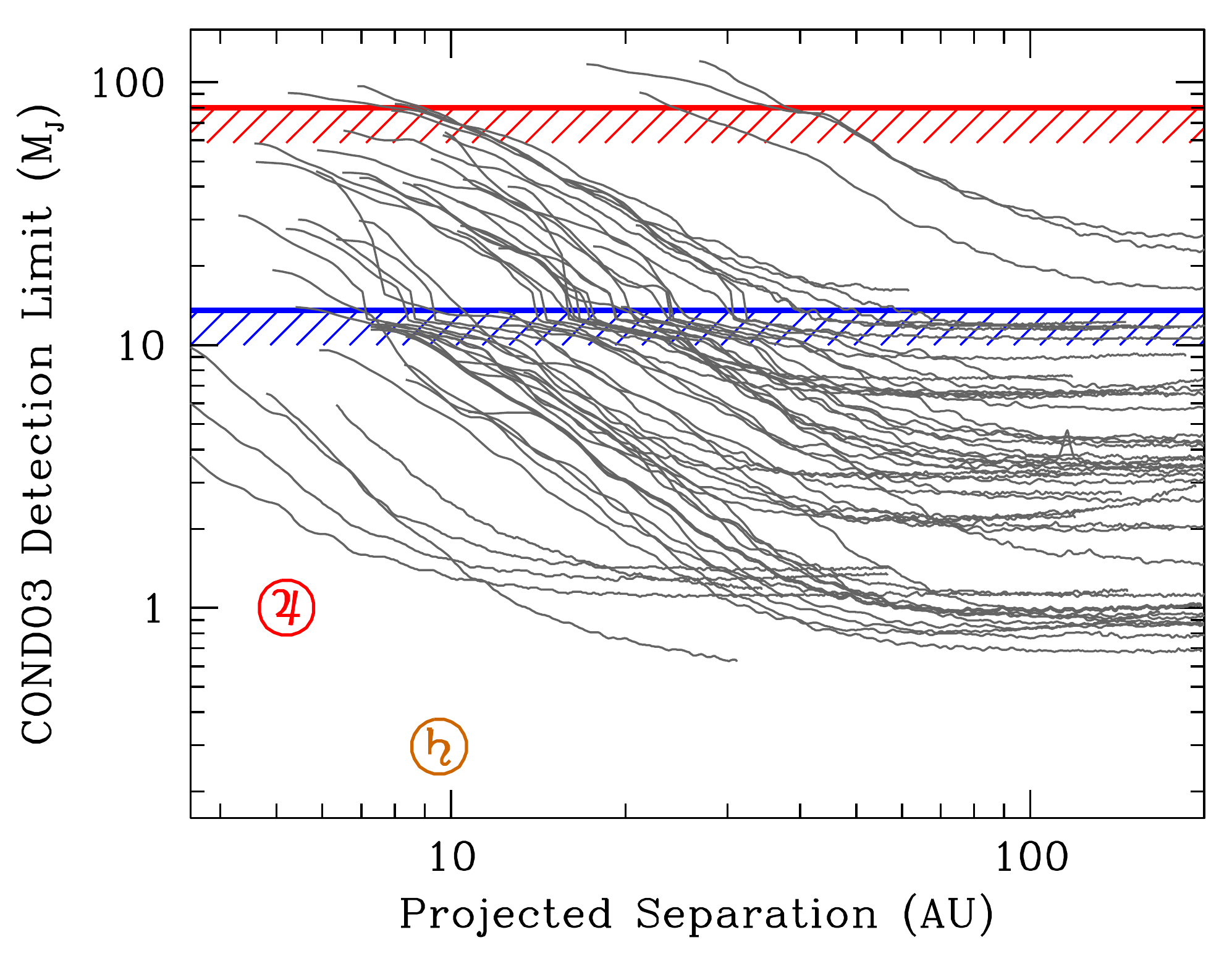}
\caption{Mass sensitivity of the SEEDS moving groups sample at the median age of the posterior probability distribution (Section \ref{sec:BayesAges}, Table \ref{tab:ages}); FK Psc, HIP 3589, HIP 6869, HIP 12925, HIP 45383, HD 95174, HIP 60661, HIP 82688, and HIP 91043 show strong artifacts from bright neighbors and have been omitted.  The COND03 models \citep{Baraffe+Chabrier+Barman+etal_2003} have been used to convert from mass to luminosity.  The red line marks the approximate stellar/brown dwarf boundary, while the blue line marks the brown dwarf/planet transition.  Jupiter and Saturn are indicated near the lower-left corner.  A thorough treatment of the statistics of the sample and its sensitivities as a function of mass will be presented in a forthcoming paper.}
\label{fig:mass_contrast}
\end{figure}

\section{Discussion}
\label{sec:discussion}

Table \ref{tab:contrast_curves} shows the $5.5\sigma$ detection limits for the SEEDS moving group targets; Figure \ref{fig:contrast_curves} plots these limits, together with 20, 50, and 80\% curves, omitting stars with contrast artifacts from nearby bright stars.  At small angular separations ($\lesssim1''$), the limiting contrast depends mostly on observing conditions, AO performance, and field rotation, with only a weak dependence on stellar brightness.  Far from the central star, SEEDS observations are read noise limited.  In this regime, limiting magnitude is a more appropriate measure than limiting contrast.  Sensitivity at these separations ($\gtrsim 2''$) depends almost exclusively on AO performance, total integration time, and integration time per frame.

The typical limiting contrast of a SEEDS observation varies from $\sim$10$^3$ at $0.\!\!''3$, to $\sim$10$^5$ at $1''$, to nearly $10^6$ at separations $\gtrsim2''$.  The limiting masses are far more variable, due to the spread in ages (and often enormous uncertainties in age) of the targets observed.  As a very crude guide to the mass sensitivity of our sample, Figure \ref{fig:mass_contrast} plots the mass detection limit as a function of projected separation around each target, assuming the median age of the posterior probability distribution (Section \ref{sec:BayesAges}, Table \ref{tab:ages}).  These sensitivities assume the COND03 exoplanet cooling models \citep{Baraffe+Chabrier+Barman+etal_2003} and neglect uncertainties in stellar age and exoplanet modeling.  We defer a full analysis of our sensitivity as a function of exoplanet mass, together with a statistical analysis of the sample and its constraints on exoplanet frequency and properties, to a forthcoming paper (Brandt et al. 2013, in preparation).  

Our sensitivity limits are competitive with other high-contrast instrumentation at other observatories, but should improve dramatically with the new extreme adaptive optics system, SCExAO, currently being commissioned at Subaru \citep{Guyon+Martinache+Clergeon+etal_2011}.  We are also exploring more minor upgrades to HiCIAO that may offer significant performance improvements.  In the Southern hemisphere, GPI \citep{Macintosh+Graham+Palmer+etal_2008} and SPHERE \citep{Beuzit+Feldt+Dohlen+etal_2008} will combine integral-field spectroscopy with high-performance adaptive optics to offer exceptional sensitivity at small angular separations.  CHARIS, an integral-field spectrograph being developed and built for the Subaru telescope, will offer similar capabilities in the Northern hemisphere \citep{McElwain+Brandt+Janson+etal_2012, Peters+Groff+Kasdin+etal_2012}.

\section{Conclusions}
\label{sec:conclusions}

We have presented high-contrast observations of 63 nearby stars in the SEEDS moving group sample.  All of the stars have been suggested to be members of coeval stellar associations.  We have reviewed each proposed association, and conclude that five associations, $\beta$ Pictoris, AB Doradus, Tucana-Horologium, Columba, and TW Hydrae are sufficiently well-defined to provide conclusive age estimates for bona-fide members.  Somewhat under half of our target sample have firm ages derived from moving group membership.

For all stars, and in particular for those without a firm moving group age, we use empirical age indicators including stellar rotation, chromospheric and coronal activity, and photospheric lithium abundance to estimate an age.  Some of these data are new observations we have acquired at the Apache Point Observatory.  The heterogeneity of our targets and their age indicators result in a wide range of constraints, with some of our targets having very precise ages and others being almost completely unconstrained.  This picture should improve as transit surveys measure photometric periods for an increasing fraction of field stars.  

We have reduced all of our observations uniformly with the recently published software ACORNS-ADI \citep{Brandt+McElwain+Turner+etal_2013} and published contrast curves for our target stars.  The contrast varies from $\sim$10$^3$ at $0.\!\!''3$ to $\sim$10$^5$ at $1''$ to $\sim$10$^6$ at $2''$; it is limited by field rotation, PSF fluctuations, and AO performance at small separations, and by AO performance and exposure time at separations $\gtrsim$$2''$.  A full analysis of our sensitivity as a function of exoplanet mass, and the constraints on exoplanet frequency and properties, is beyond the scope of this paper.  We will provide this analysis of the SEEDS moving group sample, the debris disk sample, and archival data from other surveys in a forthcoming paper.

\acknowledgements{The authors thank an anonymous referee for many helpful suggestions that substantially improved the age analysis presented in this paper.  This research is based on data collected at the Subaru Telescope, which is operated by the National Astronomical Observatories of Japan. This material is based upon work supported by the National Science Foundation Graduate Research Fellowship under grant No. DGE-0646086. Part of this research was carried out at the Jet Propulsion Laboratory, California Institute of Technology, under a contract with the National Aeronautics and Space Administration. The authors wish to recognize and acknowledge the very significant cultural role and reverence that the summit of Mauna Kea has always had within the indigenous Hawaiian community.We are most fortunate to have the opportunity to conduct observations from this mountain. This research has made use of the SIMBAD database, operated at CDS, Strasbourg, France. }

\bibliographystyle{apj_eprint}
\bibliography{seeds_refs}

\end{document}